# OGLE-2019-BLG-1470LABc: Another microlensing giant planet in a binary system?

Renkun Kuang (匡仁昆) [1,2]★ Weicheng Zang (臧伟呈) [1,]† Youn Kil Jung,[3,4]† Andrzej Udalski [5,]‡
Hongjing Yang (杨弘靖) [1,]† Shude Mao (毛淑德),[1,6]† Michael D. Albrow [7,]† Sun-Ju Chung,[3]†
Andrew Gould,[8,9]† Cheongho Han,[10]† Kyu-Ha Hwang,[3]† Yoon-Hyun Ryu,[3]† In-Gu Shin,[3]†
Yossi Shvartzvald,[11]† Jennifer C. Yee,[12]† Sang-Mok Cha,[3,13]† Dong-Jin Kim,[3]† Hyoun-Woo Kim,[3]†
Seung-Lee Kim,[3]† Chung-Uk Lee,[3]† Dong-Joo Lee,[3]† Yongseok Lee,[3,13]† Byeong-Gon Park,[3,4]† Richard
W. Pogge,[9]† Przemek Mróz [5,]‡ Jan Skowron [5,]‡ Radoslaw Poleski [5,]‡ Michał K. Szymański [5,]‡
Igor Soszyński [5,]‡ Paweł Pietrukowicz [5,]‡ Szymon Kozłowski [5,]‡ Krzysztof Ulaczyk [14,]‡ Krzysztof
A. Rybicki [5,11]‡ Patryk Iwanek [5,]‡ Marcin Wrona [5,]‡ Mariusz Gromadzki [5,]‡ Hanyue Wang (王涵
悦),[15] Shuo Huang (黄硕) [1] and Wei Zhu (祝伟) [1]

*Affiliations are listed at the end of the paper*

## ABSTRACT

We report the discovery and analysis of a candidate triple-lens single-source (3L1S) microlensing event, OGLE-2019-BLG-1470. This event was first classified as a normal binary-lens single-source (2L1S) event, but a careful 2L1S modelling showed that it needs an additional lens or source to fit the observed data. It is found that the 3L1S model provides the best fit, but the binary-lens binary-source (2L2S) model is only disfavoured by $\Delta\chi^2 \simeq 18$. All of the feasible models include a planet with planet-to-host mass-ratio $10^{-3} \lesssim q \lesssim 10^{-2}$. A Bayesian analysis based on a Galactic model indicates that the planet is super-Jovian, and the projected host-planet separation is about 3 au. Specifically, for the best-fitting 3L1S model, the two stars have masses of $M_1 = 0.57^{+0.43}_{-0.32}M_\odot$, and $M_2 = 0.18^{+0.15}_{-0.10}M_\odot$ with projected separation of $1.3^{+0.5}_{-0.4}$ au, and the planetary mass is $M_3 = 2.2^{+1.8}_{-1.3}M_{\rm Jupiter}$. For the 2L2S model, the masses of the host star and the planet are $0.55^{+0.44}_{-0.31}M_\odot$ and $4.6^{+3.7}_{-2.6}M_{\rm Jupiter}$, respectively. By investigating the properties of all known microlensing planets in binary systems, we find that all planets in binary systems published by the KMTNet survey are located inside the resonant caustics range with $q \gtrsim 2 \times 10^{-3}$, indicating the incompleteness of the KMTNet sample for planets in binary systems. Thus, planets in binary systems cannot be included in the current study of the KMTNet mass-ratio function, and a systematic search for planetary anomalies in KMTNet microlensing light curves of binary systems is needed.

## 1 INTRODUCTION

A substantial fraction of stars have one or more companion stars (e.g. Duchêne & Kraus 2013; Moe & Di Stefano 2017). The multiplicity frequency of main sequence stars is a steep monotonic function of stellar mass from ∼20 per cent for very low-mass stars (mass ≤0.1M⊙) to ≥80 per cent for high-mass stars (mass ≳16M⊙) (Duchêne & Kraus 2013). The binary fraction has the same trend (Moe & Di Stefano 2017). Both theoretical and observational studies show that stellar binarity has various effects on the protoplanet discs, e.g. driving wobbling jets with inhomogeneous accretion (Jørgensen et al. 2022) triggering misalignment in discs with the orbital plane of the binary (Zanazzi & Lai 2018; Czekala et al. 2019; Yang et al. 2020) and driving spiral arms (Dong et al. 2016). Materials can also be delivered inside the open gaps (Artymowicz & Lubow 1994) in discs to the region near the binary (Nelson & Marzari 2016; Yang et al. 2017), which can sustain supplementary materials in the circumstellar discs and thus facilitate the formation of planets orbiting one of the stars in binary systems. There are other dynamical influences from the companion. For example, the companion perturbs the protoplanetary disc, leading to a non-axisymmetric disc with non-zero eccentricity (e.g. Kley, Papaloizou & Ogilvie 2008). The eccentric disc together with the companion forms a complex environment and regulates the dynamics of the planetesimals, thus influencing planet formation (Thébault, Marzari & Scholl 2008; Marzari et al. 2013; Rafikov 2013; Rafikov & Silsbee 2015; Silsbee & Rafikov 2015). Due to complicated formation scenarios, more insightful studies on the planet formation in binary systems are still needed. The increasingly growing exoplanet census offers opportunities to reassess planet formation theories.

To date, there are about 5000[1] confirmed exoplanets and about 400 exoplanets have been confirmed in binaries with >90 per cent of these detected around FGK stars by the transit method (e.g. Kepler 16 b, Doyle et al. 2011, and Kepler 47 b, Orosz et al. 2012) or the radial velocity method (e.g. HD 147513 b, Mayor et al. 2004, and 11 Com b, Liu et al. 2008). As the planet-in-binary sample size grows, studies on comparing the characteristics of planets in binary systems with planets around single stars have been done (e.g. Roell et al. 2012). Using publicly available Kepler data, Armstrong et al. (2014) found that the occurrence rate of circumbinary planets may be consistent with or higher than that of planets orbiting a single star. Besides, with Keck II high resolution imaging for 382 Kepler Objects of Interest, Kraus et al. (2016) found that the planet occurrence rate in close binaries (with projected separation ≲50 au) is about three times lower than in wide binaries or single stars. These results indicate that the binarity may have a limited effect on the formation of circumbinary planets and circumstellar planets with very wide binary separations (≳100 au). However, for circumstellar planets[2] in much closer binaries (with separation ∼20 au), the situation is different such that the planet formation efficiency would be strongly affected (lowered) by the presence of the companion star (Thebault & Haghighipour 2015).

Within the framework of gravitational collapse or core accretion, some theoretical works indicate that a close companion would inhibit planet formation. For example, by evaporating volatile materials due to internal thermal energy generation in the disc (Nelson 2000), or by increasing the eccentricity of the gas disc and the relative velocity between dust and gas, thus reducing the coagulation and the average mass of the particles (Zsom, Sándor & Dullemond 2011). The discovery of planets in close binaries implies that planet formation is a robust process, and it has triggered a great interest in testing and developing planet formation theories in such dynamically active environments (Jang-Condell 2015). The apsidal alignment of a protoplanetary disc with the binary orbit has been found to be one of the critical conditions for planetesimal growth, which allows the emergence of a dynamically quiet location in the disc (Silsbee & Rafikov 2021).

Theories on planet formation in binary systems are under rapid development. To obtain a more complete picture of how planets form in binary systems, a larger sample would be beneficial. However, currently, there are some observational biases. For example, most circumbinary planets detected by Kepler are located near the stability limit, i.e. they would be dynamically unstable if they were in a slightly closer orbit (Holman & Wiegert 1999; Ballantyne et al. 2021). This is thought to be caused by selection effects, i.e. the transit and radial velocity methods require longer time coverage to detect longer period planets.

The gravitational microlensing technique (Mao & Paczynski 1991; Gould & Loeb 1992) is complementary to other exoplanet detection methods due to its unique sensitivity for planets in binary systems (e.g. Luhn, Penny & Gaudi 2016). With the microlensing method, the planet signal is detectable either through its influence on the central caustic formed by the stellar binary, or through the planetary caustic formed by the planet. The time-scale of a typical microlensing event towards the Galactic bulge is about one month. Unlike the transit and radial velocity methods, the microlensing method does not require years of observations due to the long orbital period of the planet or the companion. For example, Bennett et al. (2016) reported the first case of microlensing circumbinary planet. In this case, the projected separation between the planet and the centre of mass is ∼40 times larger than the separation between the two stars, well beyond the stability limit. This detection indicates that circumbinary planets with stable orbits may be quite common. However to date, only seven unambiguous planets in binary systems have been discovered by microlensing. Here we restrict the stellar binary mass-ratio, $q_2 > 0.1$. These events include OGLE-2006-BLG-284 (Bennett et al. 2020), OGLE-2007-BLG-349 (Bennett et al. 2016), OGLE-2008-BLG-092 (Poleski et al. 2014), OGLE-2013-BLG-0341 (Gould et al. 2014), OGLE-2016-BLG-0613 (Han et al. 2017),[3] OGLE-2018-BLG-1700 (Han et al. 2020), and KMT-2019-BLG-1715 (Han et al. 2021a).[4]

There are two main challenges in detecting planets in binary systems via the microlensing method. First, the perturbations on a microlensing light curve from the stellar binary are often much stronger than those from a planetary companion, for which one first needs careful binary-lens single-source (2L1S) modelling to isolate the signal from the stellar binary and then search for the planetary signals. However, in many cases, modellers would lose interest in the light curves with obvious stellar-binary features. Second, the triple-lens single-source (3L1S) modelling is computationally much more expensive than the 2L1S modelling due to a higher-dimensional parameter space and more complex image and caustics topology.

There have been several approaches to calculate 3L1S light curves. Previous analyses of 3L1S events (e.g. the first microlensing two-planet event OGLE-2006-BLG-109, Gaudi et al. 2008; Bennett et al. 2010) are mainly based on the inverse ray-shooting method (Kayser, Refsdal & Stabell 1986; Schneider & Weiss 1987), including the image centred ray-shooting method (Bennett & Rhie 1996) and the 'map-making' method (Dong et al. 2006, 2009b). In addition, Mediavilla et al. (2006), Mediavilla et al. (2011) proposed an approach based on inverse polygon mapping. In the low-magnification regime, one can use the hexadecapole approximation (Gould 2008; Pejcha & Heyrovský 2009). Recently, Kuang et al. (2021) implemented a general contour integration method (Gould & Gaucherel 1997; Dominik 1998b) for 3L1S and made this microlensing 3L1S code publicly available.[5]

In this paper, we present the first application of the code to a real 3L1S event, OGLE-2019-BLG-1470, for which the lens system is composed of a super-Jovian planet and a low-mass stellar binary. The anomaly of this event was found by the Korea Microlensing Telescope Network (KMTNet, Kim et al. 2016) AnomalyFinder (Zang et al. 2021) applied to its 2019 subprime-field sample (cadence $\Gamma < 2\,\mathrm{hr}^{-1}$), and careful 2L1S modelling conducted by H. Wang suggested that it needs an additional lens or source to fit the light curve.

The paper is structured as follows. We first introduce the observations and data reduction for this event in Section 2. We then present

---

[2] For example, γ Cephei A (Hatzes et al. 2003; Neuhäuser et al. 2007), HD 41004 A (Zucker et al. 2004), and HD 41004 A (Correia et al. 2008; Chauvin et al. 2011).

[3] For OGLE-2016-BLG-0613, there is a degenerate solution with $q_2 = 0.029 \pm 0.002$ favoured by $\Delta\chi^2 = 10$ (Han et al. 2017).
[4] We count those events with other interpretations without other competing degenerate models. Other likely candidate events of planet in binary with unresolved degenerate models include OGLE-2019-BLG-0304 (Han et al. 2021b), where the triple-lens model is favoured over the two-lens-two-source model with $\Delta\chi^2 \approx 8$. With the available data, the degeneracy cannot be securely resolved.

**Table 1.** Data used in the analysis with corresponding data reduction method (HJD′ = HJD − 2450000).

| Collaboration | Site | Name | Filter | Time coverage (HJD′) | $N_{data}$ | Reduction method | $(k, e_{min})$ |
|---|---|---|---|---|---|---|---|
| OGLE | LCO | OGLE | I | 8530.9–8763.6 | 313 | Wozniak (2000) | (1.446, 0.000) |
| KMTNet | SAAO | KMTS | I | 8584.5–8777.3 | 137 | PYSIS[1] | (1.746, 0.000) |
| KMTNet | CTIO | KMTC | I | 8546.8–8777.6 | 390 | PYSIS | (1.510, 0.010) |
| KMTNet | SSO | KMTA | I | 8563.2–8777.9 | 291 | PYSIS | (1.554, 0.000) |
| KMTNet | CTIO | KMTC | I | 8546.8–8777.6 | 390 | PYDIA[2] | – |
| KMTNet | CTIO | KMTC | V | 8542.9–8773.5 | 42 | PYDIA | – |

*Notes.* [1] Albrow et al. (2009).
[2] Albrow (2017).

the light-curve modelling process in Section 3 and the physical parameters of the lens system in Section 4. Finally, we discuss the implications derived from an examination of all microlensing planets in binary systems in Section 5.

## 2 OBSERVATIONS AND DATA REDUCTION

The microlensing event OGLE-2019-BLG-1470 at equatorial coordinates $(\alpha, \delta)_{J2000} = (18{:}07{:}47.81, -27{:}02{:}00.8)$, and Galactic coordinates $(\ell, b) = (4.1043, -3.2794)$ was announced as a candidate microlensing event by the Early Warning System (Udalski et al. 1994; Udalski 2003) of the Optical Gravitational Lensing Experiment (OGLE, Udalski, Szymański & Szymański 2015) on 2019 September 22 and independently found by the KMTNet EventFinder algorithm (Kim et al. 2018) as KMT-2019-BLG-2814 using all the data from the 2019 season. The OGLE observations were taken using its 1.3-m Warsaw Telescope equipped with a 1.4 deg² FOV mosaic CCD camera at Las Campanas Observatory (LCO) in Chile. The KMTNet data were taken using the three identical 1.6 m-telescopes equipped with 4 deg² FOV cameras at the Cerro Tololo Inter-American Observatory (CTIO) in Chile (KMTC), the South African Astronomical Observatory (SAAO) in South Africa (KMTS), and the Siding Spring Observatory (SSO) in Australia (KMTA). OGLE-2019-BLG-1470 lies in the OGLE BLG518 field and KMTNet BLG32 field with cadences of $\Gamma \sim 1$–3 night⁻¹ and $\Gamma \sim 0.4$ hr⁻¹, respectively. For both surveys, images were mainly taken in the *I*-band with occasional observations in the *V*-band for the source colour measurements. We summarize the data sets used in this work in Table 1.

The data used in the light-curve analysis were reduced using custom implementations of the difference image analysis technique (Tomaney & Crotts 1996; Alard & Lupton 1998; Bramich 2008): Wozniak (2000) for the OGLE data and PYSIS (Albrow et al. 2009) for the KMTNet data. For the KMTC data, we conduct PYDIA photometry (Albrow 2017) to measure the source colour. The *I*-band magnitude of the data has been calibrated to the standard *I*-band magnitude using the OGLE-II star catalogue (Udalski et al. 2002). Due to systematics, the photometric error bars of data estimated by photometry pipelines are often underestimated. We thus follow the method proposed by Yee et al. (2012) to adjust the error bars for each data set *i* using the formula

$$\sigma'_{i,j} = k_i \sqrt{\sigma^2_{i,j} + e^2_{i,\min}}, \tag{1}$$

where $\sigma_{i,j}$ and $\sigma'_{i,j}$ are the original and renormalized error bars in magnitudes of the *j*-th data point in the *i*-th data set. The error-bar correction parameters $k_i$ and $e_{i,\min}$ are adjusted such that $\chi^2$/dof = 1 and the cumulative sum of $\chi^2$ are approximately linear as a function of source magnification, where 'dof' is the degree of freedom. We follow the procedures above and derive the error-bar correction parameters using the best-fitting model, and other models adopt the same error-bar correction parameters. We list the data reduction methods and error-bar correction parameters for each data set in Table 1.

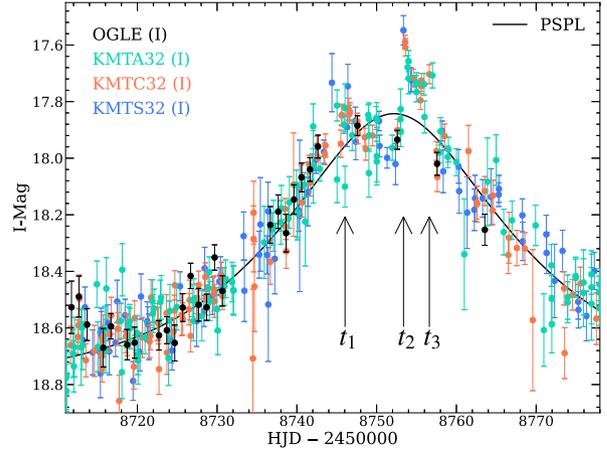

**Figure 1.** Light curve of the microlensing event OGLE-2019-BLG-1470. The dots with different colours represent the observed data from different data sets. The black solid curve is the best-fitting PSPL model. The three arrows indicate the three anomalous peaks of an otherwise PSPL model, marked as $t_1$, $t_2$, and $t_3$.

## 3 LIGHT-CURVE ANALYSIS

Fig. 1 shows the observed data of OGLE-2019-BLG-1470. The light curve exhibits three anomalous peaks relative to an otherwise normal point-source point-lens (PSPL, Paczyński 1986) light curve at HJD′ = HJD − 2450000 ∼ 8746.0, 8753.4, and 8756.6, marked as $t_1$, $t_2$, and $t_3$, respectively. The first anomaly is a smooth bump, which may be due to a cusp approach. The second and third anomalies together likely form a 'U shape', respectively corresponding to the entrance and exit of a caustic-crossing feature. Such a light curve is generally produced by a 2L1S event (e.g. Koshimoto et al. 2017), so we begin the light-curve analysis by the 2L1S modelling.

### 3.1 2L1S analysis

A static 2L1S light curve requires seven parameters $(t_0, u_0, t_E, s, q, \alpha, \rho)$ to calculate the 2L1S magnification. 'Static' means that we do not consider high-order effects such as the microlens parallax effect and the orbital motion of the lens or the source. The first three parameters of the 2L1S static model are the PSPL parameters. Of these, $t_0$ is the time of closest approach of the source to the lens centre of mass and $u_0$ is the closest distance of the source to the lens centre of mass in units of the angular Einstein radius ($\theta_E$). The third parameter $t_E$ is

**Table 2.** 2L1S lensing parameters.

| Parameter | Static | | High-order | | | |
|---|---|---|---|---|---|---|
| | | | $u_0 > 0$ | | $u_0 < 0$ | |
| | Best fit | MCMC | Best fit | MCMC | Best fit | MCMC |
| $\chi^2$/dof | 1265.25/1124 | | 1181.49/1120 | | 1181.27/1120 | |
| $t_0$ (HJD′) | 8750.870 | $8750.955^{+0.176}_{-0.172}$ | 8750.307 | $8750.265^{+0.200}_{-0.214}$ | 8750.141 | $8750.228^{+0.191}_{-0.224}$ |
| $u_0$ | 0.229 | $0.227^{+0.008}_{-0.009}$ | 0.219 | $0.228^{+0.013}_{-0.015}$ | −0.228 | $-0.229^{+0.015}_{-0.015}$ |
| $t_E$ (days) | 40.575 | $40.907^{+1.268}_{-1.106}$ | 39.576 | $39.715^{+3.377}_{-1.736}$ | 38.478 | $39.861^{+5.993}_{-1.895}$ |
| $s$ | 1.154 | $1.152^{+0.006}_{-0.006}$ | 1.171 | $1.162^{+0.010}_{-0.011}$ | 1.167 | $1.161^{+0.010}_{-0.012}$ |
| $q(10^{-3})$ | 4.456 | $4.476^{+0.392}_{-0.366}$ | 5.803 | $4.927^{+0.991}_{-1.168}$ | 5.062 | $4.691^{+1.061}_{-1.599}$ |
| $\alpha$ (rad) | 5.099 | $5.093^{+0.016}_{-0.017}$ | 5.206 | $5.173^{+0.034}_{-0.045}$ | 1.101 | $1.115^{+0.065}_{-0.038}$ |
| $\rho(10^{-3})$ | 7.56 | $<9.01\,(3\sigma)^2$ | 8.731 | $7.948^{+1.101}_{-1.222}$ | 8.271 | $7.655^{+1.106}_{-1.299}$ |
| $\pi_{E, N}$ | – | – | −2.233 | $-2.004^{+2.306}_{-0.706}$ | 2.190 | $1.841^{+0.821}_{-3.198}$ |
| $\pi_{E, E}$ | – | – | −0.716 | $-0.769^{+0.161}_{-0.183}$ | −0.857 | $-0.862^{+0.144}_{-0.173}$ |
| ds/dt(yr$^{-1}$) | – | – | −1.233 | $-0.489^{+0.788}_{-0.697}$ | −0.697 | $-0.354^{+0.365}_{-0.782}$ |
| dα/dt(yr$^{-1}$) | – | – | −2.692 | $-0.996^{+2.488}_{-1.728}$ | −0.137 | $1.048^{+1.650}_{-2.656}$ |
| $f_{S,OGLE}$ [1] | 0.194 | $0.192^{+0.009}_{-0.009}$ | 0.184 | $0.192^{+0.012}_{-0.011}$ | 0.192 | $0.192^{+0.013}_{-0.013}$ |
| $f_{B,OGLE}$ | 0.252 | $0.254^{+0.009}_{-0.009}$ | 0.267 | $0.259^{+0.015}_{-0.012}$ | 0.259 | $0.258^{+0.014}_{-0.013}$ |

*Notes.* [1] The flux is on an 18th magnitude scale, e.g. $I_S = 18 - 2.5\log(f_S)$. The reported ($f_{S,OGLE}$, $f_{B,OGLE}$) values have been calibrated to the standard *I*-band magnitude using the OGLE-III star catalogue (Udalski et al. 2002).
[2] $3\sigma$ means $\Delta\chi^2 = 9$ compared to the best-fitting $\rho$ value.

the Einstein radius crossing time, which is defined as

$$t_E = \frac{\theta_E}{\mu_{rel}}; \qquad \theta_E = \sqrt{\kappa M_L \pi_{rel}}; \qquad \kappa = \frac{4G}{c^2 au} \simeq 8.144 \frac{mas}{M_\odot}, \quad (2)$$

where $M_L$ is the lens mass and ($\pi_{rel}$, $\mu_{rel}$) are the lens-source relative (parallax, proper motion), $\mu_{rel}$ is the magnitude of the vector $\mu_{rel}$. $G$, and $c$ are the gravitational constant and the speed of light, respectively. The three additional parameters ($s$, $q$, $\alpha$) define the binary ($M_1$ and $M_2$) geometry: $s$ is the binary separation in units of $\theta_E$, $q$ is the binary mass ratio, and $\alpha$ is the angle between the source trajectory and the binary-lens axis. The last parameter, $\rho$, represents the angular source radius normalized by $\theta_E$, and it is needed to describe finite-source effects (Gould 1994; Nemiroff & Wickramasinghe 1994; Witt & Mao 1994) in caustic-crossing and/or cusp-approach features. Besides, for each data set $i$, we introduce two linear parameters ($f_{S,i}$, $f_{B,i}$) to represent the source flux and any blended flux. We use the advanced contour integration code (Bozza 2010; Bozza et al. 2018) VBBinaryLensing[6] to calculate the 2L1S magnification at any time $t$.

The static 2L1S modelling includes two steps. First, we conduct a grid search in the parameter space (log $s$, log $q$, $\alpha$, $\rho$) to find the local minima, which consists of 41 values of log $s$ equally spaced between −1.0 and 1.0, 61 values of log $q$ equally spaced between −6.0 and 0.0, 20 values equally spaced between $0° \leq \alpha < 360°$, and five values of log $\rho$ equally spaced between −3.5 and −1.5. For each grid point, we explore the parameter space of ($t_0$, $u_0$, $t_E$) with the Markov Chain Monte Carlo (MCMC) method by using the emcee ensemble sampler (Foreman-Mackey et al. 2013). We choose the sample with the minimum $\chi^2$ in the MCMC chain, and further refine it with the Nelder–Mead simplex algorithm[7] (Nelder & Mead 1965; Gao & Han 2012). The $\chi^2$ improvement with the Nelder–Mead algorithm relative to the best-fit sample in the MCMC chain is ≲1, and all the best-fitting parameters are inside the $1\sigma$ credible levels.

[6] http://www.fisica.unisa.it/GravitationAstrophysics/VBBinaryLensing.htm
[7] Throughout the paper, this is done every time we run an MCMC sampling.

Second, for the local minima identified by the grid search, we refine the solution by allowing all seven parameters to vary. We show the parameters and error bars of the best-fitting model in Table 2. Note that, as recommended by Hogg & Foreman-Mackey (2018), we use the median of the MCMC chain as the measurement value, and the 16 per cent, 84 per cent quantiles as the lower and upper $1\sigma$ error bars.

The best-fitting static 2L1S model and its residuals from the observed data are shown in Fig. 2. This model predicts a magnification higher than the observed data during $8720 < HJD' < 8734$ and $8747 < HJD' < 8752$ while during $8763 < HJD' < 8772$, the model magnification is not high enough to explain the data. In addition to test the distribution of normalized residuals against a standard normal distribution (with mean 0 and standard deviation 1), Fig. 3 shows the Quantile–Quantile (Q–Q) plot (Wilk & Gnanadesikan 1968) generated with the normalized residuals during $8720 < HJD' < 8780$. The upper left-hand panel shows the Q–Q plot of the static 2L1S model. The quantiles calculated with normalized residuals of this model obviously deviate from the standard normal distribution at both the low and high ends. We also conduct a two-sample Anderson–Darling test (Scholz & Stephens 1987) with the normalized residuals against the standard normal distribution. The null hypothesis that the residuals follow the standard normal distribution can be rejected at the 0.5 per cent significance level. We label the value of the Anderson–Darling test statistic in Fig. 3, and show the critical values for different significance levels in the caption. We are thus driven to investigate whether the residuals can be fitted by high-order effects.

The first high-order effect is the microlensing parallax effect (Gould 2000), which is due to the orbital acceleration of Earth (observer). We parametrize the effect by two parameters, $\pi_{E,N}$ and $\pi_{E,E}$, the north and east components of the microlensing parallax vector $\pi_E$ in equatorial coordinates,

$$\pi_E \equiv \frac{\pi_{rel}}{\theta_E} \frac{\mu_{rel}}{\mu_{rel}}. \qquad (3)$$

The second effect is the lens orbital motion (Dominik 1998a; Batista et al. 2011; Skowron et al. 2011), which is described by two parameters, $\mathrm{d}s/\mathrm{d}t$ and $\mathrm{d}\alpha/\mathrm{d}t$, the rates of instantaneous changes at $t_0$ of the binary separation and orientation. We find that the orbital motion effect is poorly constrained and thus restrict the MCMC trials to $\beta < 1.0$, where $\beta$ is the absolute value of the ratio of transverse kinetic to potential energy (An et al. 2002; Dong et al. 2009a),

$$\beta \equiv \left| \frac{\text{KE}_\perp}{\text{PE}_\perp} \right| = \frac{\kappa \text{M}_\odot \text{yr}^2}{8\pi^2} \frac{\pi_\text{E}}{\theta_\text{E}} \gamma^2 \left( \frac{s}{\pi_\text{E} + \pi_\text{S}/\theta_\text{E}} \right)^3;$$
$$\gamma \equiv \left( \frac{\mathrm{d}s/\mathrm{d}t}{s}, \frac{\mathrm{d}\alpha}{\mathrm{d}t} \right), \tag{4}$$

where we adopt $\pi_\text{S} \approx 0.13$ mas for parallax of the source based on the mean distance to red giant stars in the direction of this event (Nataf et al. 2013). We also consider the $u_0 > 0$ and $u_0 < 0$ solutions to the 'ecliptic degeneracy' (Jiang et al. 2004; Poindexter et al. 2005).

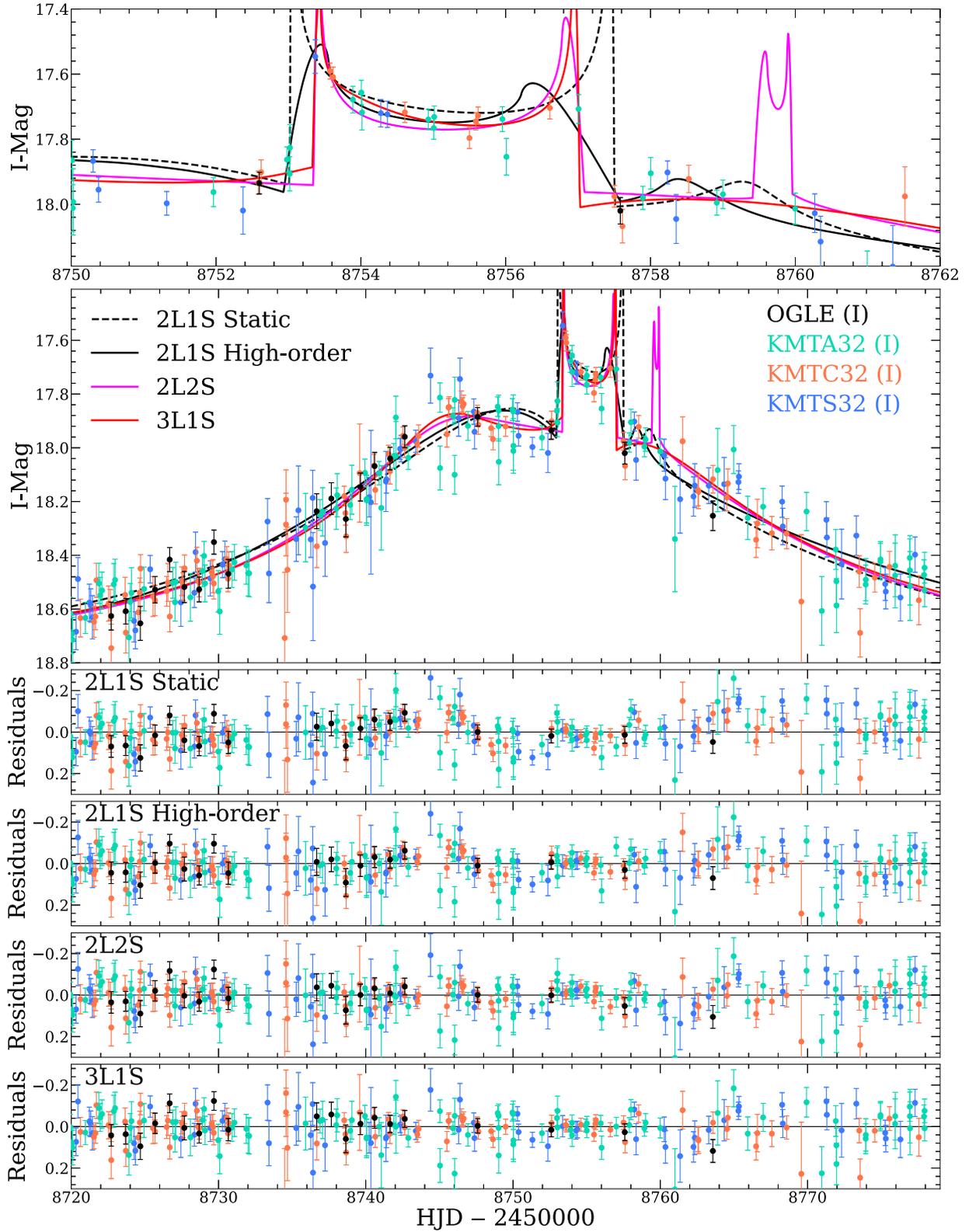

**Figure 2.** The observed data together with four models and their residuals. In the upper two panels, the dashed and solid black lines represent the best-fitting 2L1S models without and with high-order effects, respectively. The solid magenta and red lines show the best-fitting 2L2S and 3L1S models. The lower four panels show the residuals from each model.

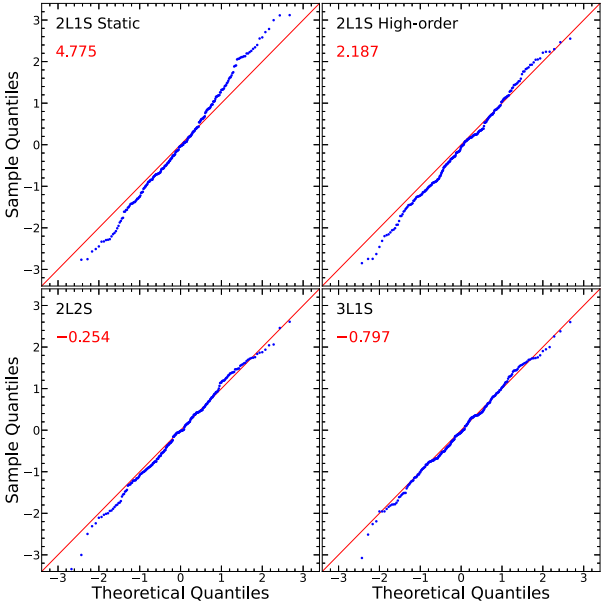

**Figure 3.** The Quantile–Quantile plots and the Anderson–Darling test results for different models. The sample quantiles are calculated with the normalized residuals during $8720 < \text{HJD}' < 8780$ (263 data points in total). The theoretical quantiles are calculated for a standard normal distribution. The red lines show the equality lines. The labelled red value at each panel shows the value of the normalized two-sample Anderson–Darling test statistic. The first sample contains 263 normalized residuals, while the second sample contains $10^7$ random values drawn from the standard normal distribution. The critical values for significance levels (25 per cent, 10 per cent, 5 per cent, 2.5 per cent, 1 per cent, 0.5 per cent, 0.1 per cent) are (0.325, 1.226, 1.961, 2.718, 3.752, 4.592, 6.546), respectively. For the '2L1S Static' model, the Anderson–Darling test statistic 4.775 is larger than the critical value for 0.5 per cent (4.592), but smaller than the critical value for 0.1 per cent (6.546). So the null hypothesis that the normalized residuals of the '2L1S Static' model follow the standard normal distribution can be rejected at the 0.5 per cent significance level. For the 2L2S and 3L1S models, the null hypothesis cannot be rejected. We used the Python packages `statsmodels` (Seabold & Perktold 2010) and `scipy` (Virtanen et al. 2020).

The parameters from the MCMC and the best-fitting model are shown in Table 2 and Fig. 2, respectively. We find that the inclusion of high-order effects improves the fit by $\Delta\chi^2 \simeq 80$. However, there are still unexplained features in this model. For example, the smooth bump in the observed data peaks at around $\text{HJD}' = 8746$, while both '2L1S Static' and '2L1S High-order' models peak at $\text{HJD}' \sim 8750$, so they predict a higher magnification than what is observed during $8747 < \text{HJD}' < 8752$. The upper right-hand panel of Fig. 3 shows the Q–Q plot of this model, in which the sample quantiles agree better with the theoretical quantiles than the static 2L1S model, but still deviate from the standard normal distribution at the low end. From the Anderson–Darling test, the null hypothesis that the residuals follow the standard normal distribution can be rejected at the 5 per cent significance level. Moreover, according to the analysis of Section 4, the $\theta_\text{E}$ and $\pi_\text{E}$ values of this model indicate a $M_\text{L} \lesssim 15 M_J$, $\mu_\text{rel} \sim 0.8$ mas yr$^{-1}$ lens system in the Galactic disc. The low lens-source relative proper motion is of fairly low probability in the Galactic dynamical model (see fig. 2 of Zhu et al. 2017), and the very low-mass lens would be rare in the context of standard stellar mass functions (e.g. Kroupa 2001). Hence, we consider models involving four objects, i.e. 3L1S (Section 3.2) and binary-lens binary-source (2L2S) (Section 3.3).

## 3.2 3L1S analysis

Relative to the 2L1S model, the 3L1S model has three additional parameters ($s_3$, $q_3$, $\psi$) to describe the third body $M_3$ (the smallest mass in our convention). Here $s_3$ and $q_3$ respectively represent the separation in units of $\theta_\text{E}$ and mass ratio between $M_1$ and $M_3$, and $\psi$ denotes the orientation angle of $M_3$ measured from the $M_1$-$M_2$ axis as seen from $M_1$. To avoid confusion, from now on we use $s_2$, $q_2$ to represent the separation and mass ratio of $M_2$ to $M_1$, i.e. $q_2 = M_2/M_1$. Due to the high-dimensional parameter space, it would be computationally expensive to conduct a grid search to explore the whole parameter space. Fortunately, the anomalies in the present case can be approximately described by the superposition of two 2L1S perturbations (Bozza 1999; Han et al. 2001; Han 2005). That is, the smooth bump at $t_1$ and the caustic-crossing feature between $t_2$ and $t_3$ can each be fitted by separate 2L1S solutions. Under this approximation, we conduct 2L1S modelling for two data subsets. In addition, we also tried other combinations of ($t_1$, $t_2$, $t_3$) and found that the resulting 3L1S models are disfavoured by $\Delta\chi^2 \gtrsim 80$ relative to the combination adopted here. For completeness, we describe those models in Section 3.2.2.

### 3.2.1 A bump versus a pair of caustic crossings

For the first data subset, we exclude the data around the caustic-crossing feature, i.e. $8752 < \text{HJD}' < 8762$. We conduct a 2L1S grid search and then refine the solutions by the MCMC with all 2L1S parameters free. We restrict the blended flux $f_{\text{B, OGLE}} > -0.2$ to exclude solutions with severe negative blended flux. We find five solutions and designate them as 'Binary A' to 'Binary E', respectively. Their parameters are presented in Table 3, and their model light curves are shown in Fig. 4. Based on the caustic geometries and source trajectories shown in Fig. 5, we find that the solutions 'Binary A' to 'Binary D' can be regarded as two pairs of solutions with $s_2 > 1$ and $s_2 < 1$. For the solution 'Binary E', its source trajectory is almost parallel to the binary axis, and such a geometry has been discovered in several previous cases (e.g. Han et al. 2017; Zhang et al. 2020). For the second data subset, we exclude the data around the smooth bump, i.e. $8735 < \text{HJD}' < 8752$. A planetary model with ($\log s_3$, $\log q_3$) $\simeq (0.05, -2.4)$ can explain this data subset, and there is no degenerate solution.

We obtain the initial parameters of 3L1S models by combining the parameters of the planetary model and each of the five binary models. Details about how to combine two 2L1S models are presented in appendix A including a detailed recipe in appendix A3. The four 3L1S solutions, which correspond to 'Binary A' to 'Binary D', can explain all the features in the observed data well with $\Delta\chi^2 \lesssim 5$ relative to the best-fitting model. We designate these 3L1S solutions as '3L1S A' to '3L1S D'. Table 4 presents the parameters of the four solutions. Their caustic structure and magnification maps together with the source trajectories are shown in Fig. 6. The model light curve of the best-fit 3L1S solution ('3L1S A') is shown in Fig. 2. The combination of the solution 'Binary E' and the planetary model cannot produce a 3L1S model that explains all the anomalies, and the resulting model from the MCMC is one of the 3L1S solutions from another combination of ($t_1$, $t_2$, $t_3$). We thus exclude this 3L1S solution which corresponds to 'Binary E'.

**Table 3.** 2L1S lensing parameters obtained from excluding a portion of the data set.

| Parameter | | | Binary | | | Planetary |
|---|---|---|---|---|---|---|
| $\chi^2$/dof | A ($s_2 < 1$) 1078.49/1082 | B ($s_2 > 1$) 1080.59/1082 | C ($s_2 < 1$) 1078.24/1082 | D ($s_2 > 1$) 1079.08/1082 | E 1072.86/1082 | 1033.06/1040 |
| $t_0$ (HJD$'$) | $8750.180^{+0.710}_{-0.731}$ | $8750.432^{+0.748}_{-2.742}$ | $8753.517^{+0.644}_{-0.563}$ | $8751.418^{+0.272}_{-0.354}$ | $8753.794^{+0.212}_{-0.217}$ | $8752.944^{+0.313}_{-0.299}$ |
| $u_0$ | $0.263^{+0.035}_{-0.038}$ | $0.297^{+0.135}_{-0.176}$ | $0.362^{+0.101}_{-0.083}$ | $0.374^{+0.060}_{-0.048}$ | $0.564^{+0.042}_{-0.048}$ | $0.203^{+0.020}_{-0.023}$ |
| $t_E$ (days) | $35.267^{+3.763}_{-3.037}$ | $32.378^{+5.376}_{-4.783}$ | $33.274^{+4.452}_{-3.780}$ | $33.578^{+3.536}_{-3.874}$ | $28.332^{+1.892}_{-1.941}$ | $44.346^{+3.326}_{-2.662}$ |
| $s_2$ (for Binary) or $s_3$ (for Planetary) | $0.469^{+0.036}_{-0.021}$ | $2.225^{+0.272}_{-0.154}$ | $0.484^{+0.021}_{-0.020}$ | $1.422^{+0.090}_{-0.101}$ | $1.161^{+0.028}_{-0.029}$ | $1.130^{+0.010}_{-0.011}$ |
| $q_2$ or $q_3$ | $0.595^{+0.362}_{-0.293}$ | $0.068^{+0.058}_{-0.025}$ | $0.905^{+1.060}_{-0.441}$ | $0.106^{+0.030}_{-0.023}$ | $0.226^{+0.055}_{-0.026}$ | $4.342^{+1.179}_{-0.762}(\times 10^{-3})$ |
| $\alpha$ (rad) | $4.081^{+0.119}_{-0.093}$ | $4.181^{+0.062}_{-0.073}$ | $2.487^{+0.163}_{-0.228}$ | $1.811^{+0.045}_{-0.046}$ | $0.124^{+0.028}_{-0.026}$ | $4.936^{+0.031}_{-0.030}$ |
| $\rho(10^{-2})$ | $<5.00(3\sigma)$ | $<5.00(3\sigma)$ | $<5.00(3\sigma)$ | $<4.99(3\sigma)$ | $<4.98(3\sigma)$ | $<1.26(3\sigma)$ |
| $f_{S,\,OGLE}$ | $0.225^{+0.037}_{-0.036}$ | $0.372^{+0.156}_{-0.099}$ | $0.310^{+0.113}_{-0.083}$ | $0.280^{+0.083}_{-0.083}$ | $0.524^{+0.077}_{-0.069}$ | $0.157^{+0.018}_{-0.020}$ |
| $f_{B,\,OGLE}$ | $0.224^{+0.036}_{-0.037}$ | $0.076^{+0.098}_{-0.155}$ | $0.139^{+0.083}_{-0.112}$ | $0.169^{+0.049}_{-0.083}$ | $-0.075^{+0.069}_{-0.077}$ | $0.291^{+0.020}_{-0.018}$ |

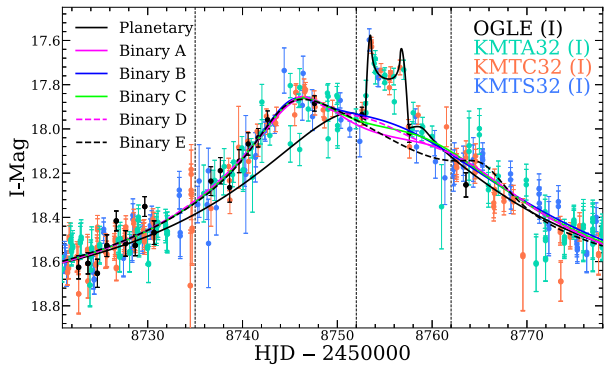

**Figure 4.** The best-fitting 2L1S model light curves obtained by excluding a portion of the data. For the planetary model (solid black), the data inside the range of $8735 <$ HJD$' < 8752$ are excluded. For the binary models, the data inside the region $8752 <$ HJD$' < 8762$ are excluded. Their corresponding caustic structure and source trajectories are shown in Fig. 5. The three vertical dashed lines correspond to HJD$' = 8735$, 8752, and 8762, respectively.

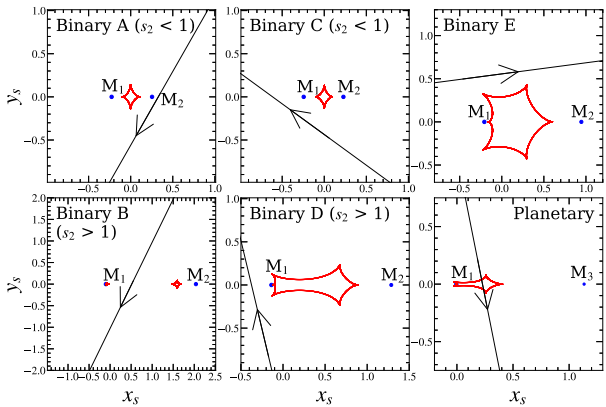

**Figure 5.** Caustic geometries of the 2L1S solutions obtained by excluding a portion of the data. Their corresponding model light curves are shown in Fig. 4.

We note that some of the reported 3L1S model parameters and uncertainties (Table 4) may be different from the 2L1S models (Table 3) they are built upon. For example, the values of $u_0$ are different for 'Binary D' in Table 3 and '3L1S D' in Table 4. This is because by adding a planetary mass into the binary models, parameters like $t_E$, $u_0$ are better constrained by the caustic crossing feature in the light curve caused by the planet mass. Other parameters would also be constrained better since parameters are correlated. Actually, after obtaining the 2L1S models for each data subset, one could further do a joint fit before 3L1S modelling. Because for each pair of 'Binary' and 'Planetary' models, they should have at least the same $t_E$ and $\rho$.

We find that the inclusion of high-order effects improve the fit by $\Delta\chi^2 \simeq 3$ and the $1\sigma$ uncertainty of parallax is $\sigma(\pi_{E,\,\|}) \simeq 0.2$ and $\sigma(\pi_{E,\,\perp}) \simeq 1.0$, where $\pi_{E,\,\|}$ and $\pi_{E,\,\perp}$ is the component of $\pi_E$ that is parallel and perpendicular with the direction of Earth's acceleration. In addition, the other 3L1S parameters are consistent with those of the static 3L1S models at $1\sigma$. Such a weak constraint on $\pi_E$ has little effect ($< 5$ per cent) on the physical parameters derived from the Bayesian analysis in Section 4.2. Hence, we adopt the static 3L1S models as the final results.

### 3.2.2 Other combinations of features

Currently, the solutions of most known 3L1S events were found by the superposition of two 2L1S perturbations. The strategy of finding 3L1S models from combinations of 2L1S models is mainly arrived at by inspecting the features of the observed light curve. If the data coverage of each anomaly is good and thus the shape of each anomaly is clear, then the combinations are relatively straightforward. For example, the two planetary signals of the 3L1S event OGLE-2012-BLG-0026 (Han et al. 2013) clearly consist of a bump and a dip, so only one combination of light curve features is feasible. However, in the present case, the coverage at the caustic crossings of the 'U shape' anomaly ($t_2$ and $t_3$) is sparse, so it raises the question of whether or not other combinations of ($t_1$, $t_2$, $t_3$) can yield a reasonable 3L1S solution.

**Table 4.** 3L1S lensing parameters together with $\theta_*$ and $\theta_E$.

| Parameter | A ($s_2 < 1$) | | B ($s_2 > 1$) | | C ($s_2 < 1$) | | D ($s_2 > 1$) | |
|---|---|---|---|---|---|---|---|---|
| | Best fit | MCMC | Best fit | MCMC | Best fit | MCMC | Best fit | MCMC |
| $\chi^2$/dof | 1121.45/1121 | | 1124.81/1121 | | 1126.78/1121 | | 1126.26/1121 | |
| $t_0$ (HJD′) | 8751.021 | $8751.032^{+0.190}_{-0.173}$ | 8741.156 | $8739.940^{+1.680}_{-2.020}$ | 8753.446 | $8753.262^{+0.212}_{-0.195}$ | 8718.708 | $8716.294^{+4.161}_{-3.698}$ |
| $u_0$ | 0.187 | $0.192^{+0.018}_{-0.011}$ | −0.049 | $-0.068^{+0.032}_{-0.036}$ | 0.295 | $0.285^{+0.016}_{-0.018}$ | 0.936 | $0.962^{+0.053}_{-0.051}$ |
| $t_E$ (days) | 42.595 | $41.725^{+1.827}_{-3.111}$ | 45.378 | $46.890^{+2.544}_{-2.113}$ | 36.243 | $36.302^{+1.340}_{-0.836}$ | 55.622 | $56.037^{+2.599}_{-1.990}$ |
| $s_2$ | 0.439 | $0.446^{+0.021}_{-0.017}$ | 2.749 | $2.786^{+0.094}_{-0.083}$ | 0.457 | $0.451^{+0.012}_{-0.013}$ | 2.700 | $2.768^{+0.094}_{-0.078}$ |
| $q_2$ | 0.359 | $0.322^{+0.049}_{-0.073}$ | 0.173 | $0.180^{+0.023}_{-0.020}$ | 0.975 | $1.019^{+0.185}_{-0.200}$ | 0.695 | $0.708^{+0.061}_{-0.065}$ |
| $\alpha$ (rad) | 3.900 | $3.890^{+0.029}_{-0.011}$ | 3.913 | $3.888^{+0.045}_{-0.037}$ | 2.416 | $2.424^{+0.021}_{-0.062}$ | 2.234 | $2.238^{+0.042}_{-0.036}$ |
| $s_3$ | 1.108 | $1.112^{+0.017}_{-0.011}$ | 1.059 | $1.053^{+0.009}_{-0.010}$ | 1.019 | $1.011^{+0.016}_{-0.020}$ | 0.870 | $0.867^{+0.014}_{-0.014}$ |
| $q_3(10^{-3})$ | 3.472 | $3.878^{+1.132}_{-0.545}$ | 4.953 | $5.149^{+0.604}_{-0.501}$ | 8.643 | $9.359^{+1.676}_{-1.400}$ | 4.275 | $3.996^{+0.391}_{-0.421}$ |
| $\psi$ (rad) | 4.993 | $4.998^{+0.034}_{-0.032}$ | 5.184 | $5.153^{+0.057}_{-0.050}$ | 4.093 | $4.087^{+0.033}_{-0.087}$ | 3.615 | $3.609^{+0.054}_{-0.055}$ |
| $\rho(10^{-2})$ | 0.121 | <1.148 (3σ) | 0.885 | <1.138 (3σ) | 0.928 | <1.321 (3σ) | 0.572 | <0.955 (3σ) |
| $f_{S, OGLE}$ | 0.158 | $0.163^{+0.021}_{-0.021}$ | 0.181 | $0.171^{+0.018}_{-0.018}$ | 0.247 | $0.242^{+0.012}_{-0.015}$ | 0.194 | $0.193^{+0.015}_{-0.019}$ |
| $f_{B, OGLE}$ | 0.290 | $0.285^{+0.010}_{-0.021}$ | 0.267 | $0.278^{+0.018}_{-0.018}$ | 0.201 | $0.207^{+0.015}_{-0.012}$ | 0.253 | $0.254^{+0.019}_{-0.016}$ |
| $\theta_*$ (μas) | 0.62 | $0.63^{+0.08}_{-0.06}$ | 0.66 | $0.64^{+0.08}_{-0.08}$ | 0.78 | $0.77^{+0.08}_{-0.08}$ | 0.69 | $0.69^{+0.08}_{-0.09}$ |
| $\theta_E$ (mas) | 0.51 | >0.055 (3σ) | 0.075 | >0.056 (3σ) | 0.084 | >0.058 (3σ) | 0.12 | >0.072 (3σ) |

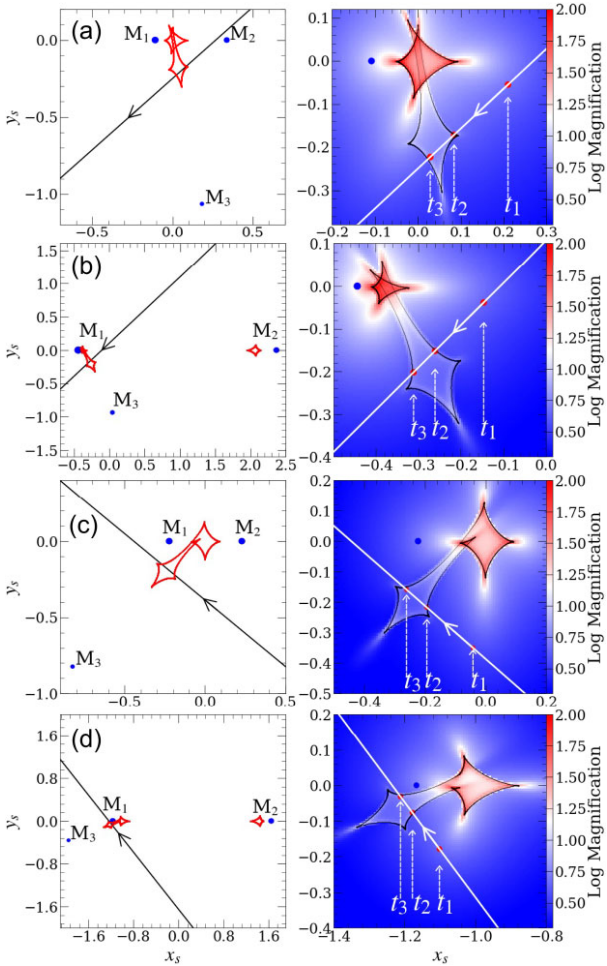

**Figure 6.** Caustic geometries of the 3L1S solutions. From top to bottom, the panels show the solutions A ($s_2 < 1$), B ($s_2 > 1$), C ($s_2 < 1$), and D ($s_2 > 1$), respectively. In each row, the left-hand panel exhibits the global view of the system, the solid black line with arrow shows the source trajectory, and the blue circles represent the locations of the lens components marked as $M_1$, $M_2$, and $M_3$. The right-hand panel displays the magnification map in which the dashed white lines with arrows indicate the times ($t_1$, $t_2$, $t_3$) of the anomalies and red dots indicate the corresponding source positions.

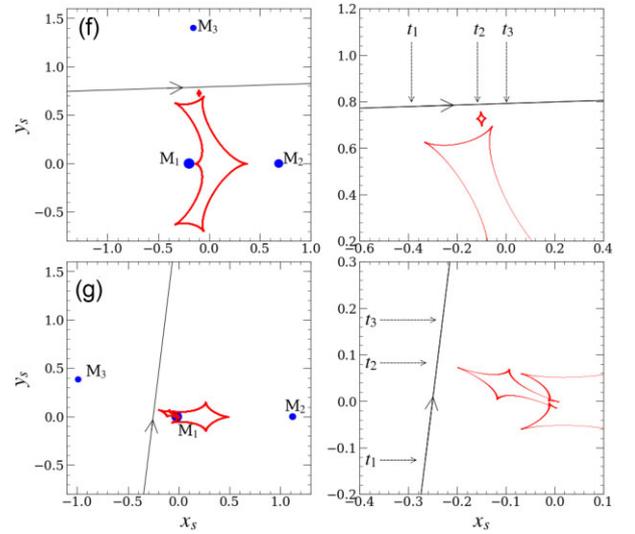

**Figure 7.** Caustic geometries of solutions '3L1S F' and '3L1S G'. The two solutions are found by assuming that the anomalies at $t_1$ and $t_3$ are caused by one 2L1S solution and the anomaly at $t_2$ alone is produced by the other 2L1S solution. The right-hand panels display a close-up of positions for the three anomalies at ($t_1$, $t_2$, $t_3$). The corresponding model light curves are shown in Fig. 8.

The first combination assumes that the anomalies at $t_1$ and $t_3$ are caused by one 2L1S solution and the anomaly at $t_2$ alone is produced by the other 2L1S solution. We first exclude the data at $t_2$, i.e. 8752 < HJD′ < 8754.5, and conduct a 2L1S grid search with the remaining data, which yields two solutions. As shown in Fig. 7, two cusp approaches with a resonant caustic produce the anomalies at $t_1$ and $t_3$. Then, we find several planetary solutions that can explain the bump feature at $t_2$. However, the 3L1S solutions from this combination are all disfavoured with $\Delta\chi^2 \gtrsim 80$ relative to the '3L1S A' solution, and these 3L1S solutions cannot fit the anomalies at $t_2$ and $t_3$. For simplicity, we only display two representatives of these 3L1S solutions (designated as '3L1S F' and '3L1S G'). Their parameters and model curves are shown in Table 5 and Fig. 8, respectively.

**Table 5.** 3L1S lensing parameters of '3L1S F' and '3L1S G'.

| Parameter | F | | G | |
|---|---|---|---|---|
| | Best fit | MCMC | Best fit | MCMC |
| $\chi^2$/dof | 1207.04/1121 | | 1200.80/1121 | |
| $t_0$ (HJD$'$) | 8755.747 | $8755.606^{+0.298}_{-0.301}$ | 8751.555 | $8751.560^{+0.136}_{-0.121}$ |
| $u_0$ | 0.775 | $0.732^{+0.073}_{-0.081}$ | 0.229 | $0.240^{+0.031}_{-0.026}$ |
| $t_E$ (days) | 27.250 | $27.471^{+1.224}_{-1.164}$ | 36.801 | $35.793^{+3.242}_{-2.825}$ |
| $s_2$ | 0.887 | $0.902^{+0.023}_{-0.022}$ | 1.163 | $1.154^{+0.041}_{-0.043}$ |
| $q_2$ | 0.274 | $0.242^{+0.063}_{-0.055}$ | 1.802 $(\times 10^{-2})$ | $1.766^{+0.277}_{-0.230}(\times 10^{-2})$ |
| $\alpha$ (rad) | 0.040 | $0.052^{+0.036}_{-0.029}$ | 1.444 | $1.451^{+0.015}_{-0.015}$ |
| $s_3$ | 1.388 | $1.339^{+0.062}_{-0.062}$ | 1.037 | $1.015^{+0.046}_{-0.033}$ |
| $q_3(10^{-3})$ | 0.793 | $1.267^{+0.518}_{-0.371}$ | 0.680 | $0.894^{+0.478}_{-0.353}$ |
| $\psi$ (rad) | 1.549 | $1.555^{+0.049}_{-0.049}$ | 2.753 | $2.768^{+0.027}_{-0.028}$ |
| $\rho(10^{-2})$ | 1.822 | $<2.905(3\sigma)$ | 1.196 | $<1.828(3\sigma)$ |
| $f_{S,\,OGLE}$ | 0.761 | $0.582^{+0.041}_{-0.074}$ | 0.203 | $0.214^{+0.033}_{-0.029}$ |
| $f_{B,\,OGLE}$ | −0.312 | $-0.133^{+0.074}_{-0.041}$ | 0.246 | $0.235^{+0.033}_{-0.033}$ |

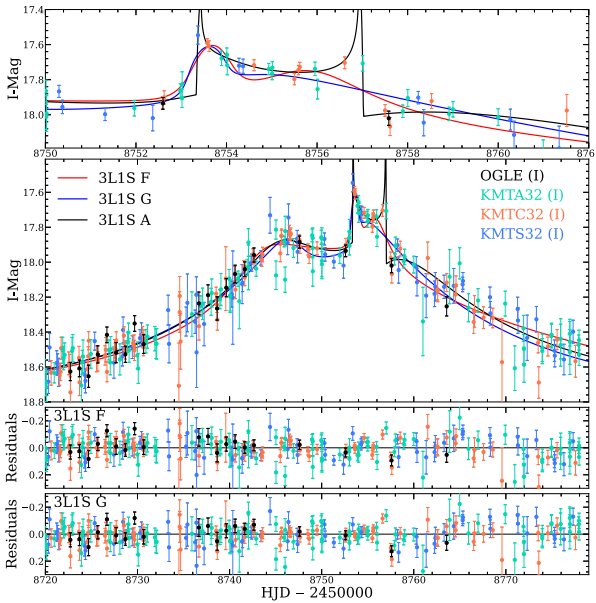

**Figure 8.** Comparison between '3L1S F' and '3L1S G' solutions with the best-fit 3L1S solution ('3L1S A'). The two solutions are disfavoured by $\Delta\chi^2 \sim 80$.

The second combination assumes that the anomalies at $t_1$ and $t_2$ are produced by one 2L1S system. We exclude data around $t_3$, i.e. $8754.5 < \mathrm{HJD}' < 8757.5$. However, we cannot find any 2L1S model that can explain the remaining data. Thus, this approach is infeasible.

Although our attempts above did not yield a new competitive 3L1S solution and the anomalies at $t_2$ and $t_3$ together indeed form a 'U shape', one may need to be cautious about the strategies of 2L1S combinations with numerous 3L1S events detected by the ongoing KMTNet survey and the Nancy Grace Roman Space Telescope (Spergel et al. 2015; Penny et al. 2019) in the future.

We note that in finding 3L1S solutions, we tried only the supposition method and have not conducted a thorough grid search over the 3L1S parameter space. We believe there is little chance for this event to have other 3L1S solutions. We have already tested all combinations of anomaly features in the above sections and conducted thorough grid searches for both the binary and planetary 2L1S models that can produce the anomaly features in the observed light curve. In this event, the planet ($M_3$) with mass-ratio $q_3 \sim 10^{-3}$ has little effect on the caustic of the binary ($M_1$ and $M_2$). So the binary-superposition method would be valid for this event.

### 3.3 2L2S analysis

There have been several events with plausible 3L1S planetary solutions that proved to be 2L2S events (e.g. Jung et al. 2017) or have competitive 2L2S solutions (e.g. Suzuki et al. 2018). The total magnification of a 2L2S model is the superposition of two 2L1S models involved with the individual source stars,

$$A_\lambda = \frac{A_1 f_{1,\lambda} + A_2 f_{2,\lambda}}{f_{1,\lambda} + f_{2,\lambda}} = \frac{A_1 + f_{\mathrm{ratio},\lambda} A_2}{1 + f_{\mathrm{ratio},\lambda}}; \quad f_{\mathrm{ratio},\lambda} \equiv \frac{f_{2,\lambda}}{f_{1,\lambda}}, \quad (5)$$

where $A_\lambda$ is the total magnification, and $f_{i,\lambda}$ is the baseline flux at wavelength $\lambda$ of each source with $i = 1$ and 2 corresponding to the primary and the secondary sources, respectively. To include the second source, we require four additional parameters, $(t_{0,2}, u_{0,2}, \rho_2, f_{\mathrm{ratio},1})$ (Hwang et al. 2013). $t_{0,2}$ is the time at which the second source is closest to the centre of mass of the lens, $u_{0,2}$ is the lens–source separation at that time, $\rho_2$ is the normalized radius of the second source, and $f_{\mathrm{ratio},1}$ is the source flux ratio in the $I$-band. We use the best-fitting parameters of the static 2L1S model as the initial parameters of $(t_{0,1}, u_{0,1}, t_E, s, q, \alpha, \rho_1)$, and use the MCMC method to generate samples from the posterior distribution, and search for the best-fitting 2L2S model with the Nelder–Mead algorithm.

Table 6 lists the parameters of the 2L2S model, Fig. 2 shows its model curve and its residuals. It is found that the 2L2S model provides a better fit than the 2L1S model by $\Delta\chi^2 = 126$. The goodness of fit improved because the second source is relatively 'delayed' compared with the first source. Fig. 9 shows the trajectories of the two sources. At time $t_1$, the first source is nearly at its closest approach to the primary lens, thus would cause a strong bump, as is the case in the static 2L1S model. In the mean time, the second source is located at the low magnification region between two spikes, which causes the total magnification at time $t_1$ to be lower and improves the goodness of fit. The 2L2S model has a $\Delta\chi^2 \simeq 18$ compared to the 3L1S model, the $\Delta\chi^2$ is mainly accumulated at time around the caustic crossing region. See Fig. 10 for the cumulative $\Delta\chi^2$ relative to the 3L1S model as a function of time. However, we cannot firmly resolve the degeneracy between this model and the 3L1S models with the current data. The lower left-hand and lower right-hand panels of Fig. 3 show the Q–Q plots of the 2L2S and the 3L1S models, respectively. For both models, the null hypothesis that the residuals follow the standard normal distribution cannot be firmly rejected. Including the parallax effect in the 2L2S model only improves the $\chi^2$ by 1.7, so we adopt the static 2L2S model as the result for simplicity.

**Table 6.** 2L2S lensing parameters together with $\theta_*$ and $\theta_E$.

| Parameter | Best fit | MCMC |
|---|---|---|
| $\chi^2$/dof | 1139.25/1120 | |
| $t_{0,1}$ (HJD$'$) | 8746.251 | $8746.510^{+0.375}_{-0.322}$ |
| $t_{0,2}$ (HJD$'$) | 8757.144 | $8757.096^{+0.445}_{-0.461}$ |
| $u_{0,1}$ | 0.150 | $0.165^{+0.012}_{-0.013}$ |
| $u_{0,2}$ | $-0.238$ | $-0.260^{+0.034}_{-0.042}$ |
| $t_E$ (days) | 36.729 | $35.544^{+1.959}_{-2.130}$ |
| $s$ | 1.215 | $1.218^{+0.017}_{-0.015}$ |
| $q(10^{-3})$ | 7.558 | $8.083^{+1.154}_{-1.128}$ |
| $\alpha$ (rad) | 5.696 | $5.654^{+0.039}_{-0.048}$ |
| $f_{ratio,I}$ | 1.600 | $1.533^{+0.405}_{-0.320}$ |
| $\rho_1(10^{-2})$ | 0.092 | $<0.665\,(3\sigma)$ |
| $\rho_2(10^{-2})$ | 0.039 | $<21.3\,(3\sigma)$ |
| $f_{S,\,OGLE}$ | 0.194 | $0.210^{+0.031}_{-0.023}$ |
| $f_{B,\,OGLE}$ | 0.255 | $0.239^{+0.023}_{-0.031}$ |
| $\theta_*$ ($\mu$as) | 0.69 | $0.72^{+0.10}_{-0.09}$ |
| $\theta_E$ (mas) | 0.75 | $>0.11$ |

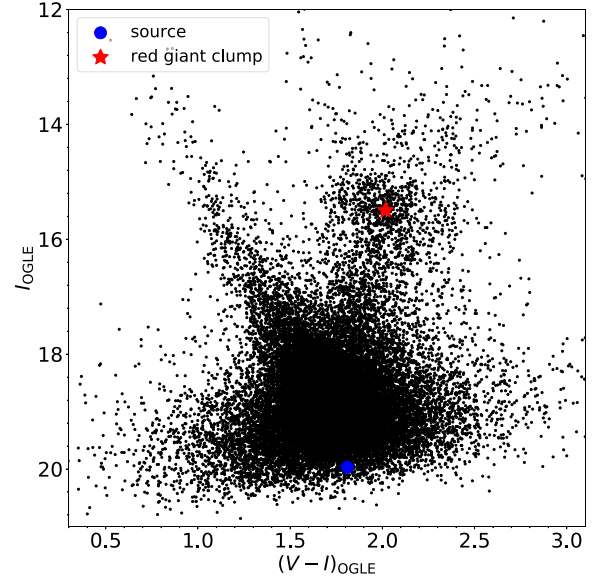

**Figure 11.** Colour–magnitude diagram (CMD) for field stars (black dots) within 180 arcsec centred on OGLE-2019-BLG-1470 using the OGLE-II catalogue stars (Udalski et al. 2002). The red asterisk indicates the centroid of the red giant clump, and the blue dot represents the microlensing source star.

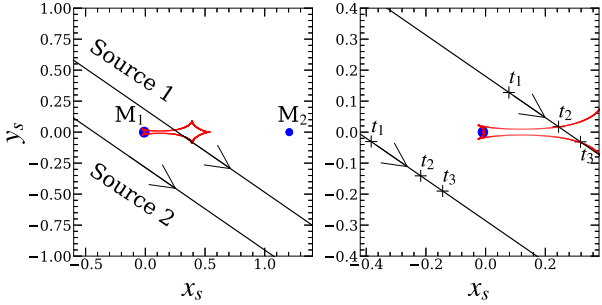

**Figure 9.** Caustic geometry and source trajectories of the 2L2S model. The red curve shows the caustic. The black lines with arrows show the trajectories of the two sources. In the right-hand panel, we mark the source positions at time $t_1$, $t_2$, and $t_3$ with crosses.

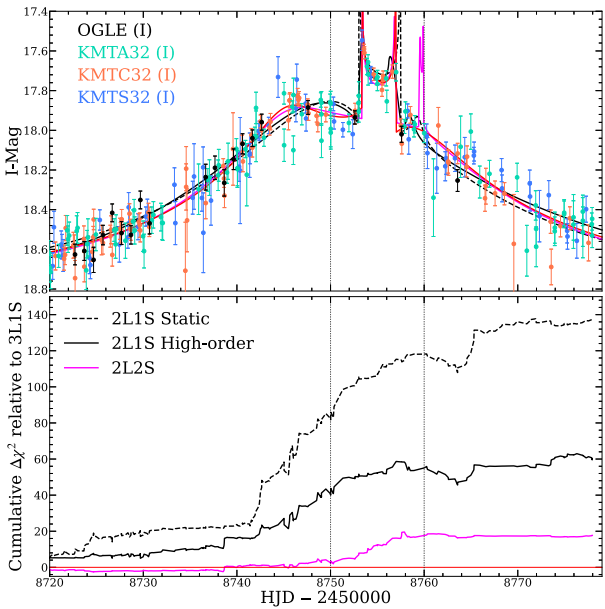

**Figure 10.** Cumulative $\Delta\chi^2$ of different models relative to the best-fitting 3L1S model.

# 4 PHYSICAL LENS PROPERTIES

For a lensing object, the mass ($M_L$) and distance ($D_L$) of the lens system are related to the angular Einstein radius $\theta_E$ and microlensing parallax $\pi_E$ by (Gould 2000)

$$M_L = \frac{\theta_E}{\kappa\pi_E}; \qquad D_L = \frac{au}{\pi_E\theta_E + \pi_S}. \qquad (6)$$

In the present case, because neither $\theta_E$ nor $\pi_E$ is unambiguously measured, we conduct a Bayesian analysis by incorporating priors from a Galactic model to estimate the physical parameters of the lens system. Before the Bayesian procedures, we estimate the angular source radius $\theta_*$ by a colour magnitude diagram (CMD, Yoo et al. 2004) analysis and then obtain the constraint on $\theta_E$ through $\theta_E = \theta_*/\rho$.

## 4.1 Colour magnitude diagram and angular source radius

We construct a $V - I$ versus $I$ colour-magnitude diagram (CMD) using the OGLE-II catalogue (Udalski et al. 2002) for field stars within 180 arcsec centred on the event. The CMD is shown in Fig. 11. We find the centroid of the red giant clump of $(V - I, I)_{cl} = (2.02 \pm 0.01, 15.50 \pm 0.02)$ and adopt $(V - I, I)_{cl,0} = (1.06, 14.33)$ (Bensby et al. 2013; Nataf et al. 2013) as the intrinsic colour and de-reddened magnitude of the red giant clump. For the source colour, which is model independent, we obtain $(V − I)_S = 1.81 \pm 0.07$ from a regression of the KMTC $V$ versus $I$ flux as the lensing magnification changes and a calibration to the OGLE-II magnitudes. Because the four 3L1S solutions have different source fluxes, we begin with the angular source radius of solution '3L1S A', $\theta_{*,A}$, using $I_S = 19.97^{+0.14}_{-0.07}$. We obtain the source de-reddened colour and magnitude as

$$
\begin{aligned}
(V − I, I)_{S,0} &= (V − I, I)_S − (V − I, I)_{cl} + (V − I, I)_{cl,0} \\
&= (0.85 \pm 0.08, 18.80^{+0.15}_{-0.08}).
\end{aligned} \tag{7}
$$

**Table 7.** Physical parameters of the lens system from a Bayesian analysis.

| Parameter | 3L1S | | | | 2L2S |
|---|---|---|---|---|---|
| | A ($s_2 < 1$) | B ($s_2 > 1$) | C ($s_2 < 1$) | D ($s_2 > 1$) | |
| $M_1$ ($M_\odot$) | $0.57^{+0.43}_{-0.32}$ | $0.58^{+0.43}_{-0.32}$ | $0.56^{+0.44}_{-0.32}$ | $0.60^{+0.43}_{-0.32}$ | $0.55^{+0.44}_{-0.31}$ |
| $M_2$ ($M_\odot$) | $0.18^{+0.15}_{-0.10}$ | $0.10^{+0.08}_{-0.06}$ | $0.55^{+0.45}_{-0.32}$ | $0.42^{+0.30}_{-0.23}$ | – |
| $M_3$ ($M_J$) | $2.2^{+1.8}_{-1.3}$ | $3.1^{+2.3}_{-1.7}$ | $5.4^{+4.3}_{-3.1}$ | $2.5^{+1.8}_{-1.4}$ | $4.6^{+3.7}_{-2.6}$ |
| $r_{\perp, 2}$ (au) | $1.3^{+0.5}_{-0.5}$ | $7.8^{+2.8}_{-2.8}$ | $1.6^{+0.6}_{-0.6}$ | $9.5^{+5.3}_{-3.4}$ | – |
| $r_{\perp, 3}$ (au) | $3.2^{+1.2}_{-1.2}$ | $2.9^{+1.1}_{-1.1}$ | $3.5^{+1.3}_{-1.3}$ | $3.0^{+1.1}_{-1.1}$ | $3.2^{+1.2}_{-1.2}$ |
| $D_L$ (kpc) | $5.9^{+1.2}_{-2.7}$ | $5.6^{+1.4}_{-2.7}$ | $6.1^{+1.1}_{-2.5}$ | $5.1^{+1.9}_{-2.5}$ | $6.1^{+1.0}_{-2.5}$ |
| $\mu_{\rm rel}$ (mas yr$^{-1}$) | $4.9^{+3.6}_{-2.0}$ | $4.3^{+3.4}_{-1.8}$ | $6.4^{+4.4}_{-2.6}$ | $4.8^{+4.1}_{-2.2}$ | $4.6^{+3.1}_{-1.8}$ |

Using the colour/surface-brightness relation for dwarfs and sub-giants of Adams, Boyajian & von Braun (2018), we obtain $\theta_{*,A} = 0.63^{+0.08}_{-0.06}$ μas. Then, for any model with source magnitude $I_S$, one can infer $\theta_* = \theta_{*,A} \times 10^{-0.2(I_S − 19.97)}$. We list $\theta_*$ of each 3L1S solutions in Table 4.

### 4.2 Bayesian analysis

The Galactic model used for the Bayesian analysis has three parts: the mass function of the lens, the stellar number density profile and the dynamical distributions. For the lens mass function, we apply the initial mass function (IMF) of Kroupa (2001) and add a 1.3$M_\odot$ and 1.1$M_\odot$ upper-end truncation for the disc and the bulge lenses, respectively Zhu et al. (2017). For the stellar number density, we choose the models used by Yang et al. (2021). For the disc velocity distribution, we use the 'Model C' of Yang et al. (2021), which is dynamically self-consistent with the density profile. For the bulge dynamical distributions, we adopt the model used by Zhu et al. (2017) and assume that the bulge stars have a zero mean velocity and 120 km s$^{-1}$ velocity dispersion along each direction.

We create a sample of $10^8$ simulated events drawn from the Galactic model. For each simulated event, $i$, whose parameters consist of $t_{E,i}$, $\mu_{rel,i}$, and $t_{E,i}$, we weight it by

$$
\omega_{\rm Gal,i} = \theta_{E,i} \times \mu_{rel,i} \times \mathcal{L}(t_{E,i}) \mathcal{L}(\theta_{E,i}). \tag{8}
$$

where $\mathcal{L}(t_{E,i})$ is the probability of $t_{E,i}$ given the error distributions of $t_E$ derived from the MCMC chain, and $\mathcal{L}(\theta_{E,i})$ is the probability of $\theta_{E,i}$. To derive the probability distribution of $\theta_E$, we first draw the probability distribution of $\rho$ by the lower envelope of $\chi^2$ versus $\rho$ diagram from MCMC. See fig. 6 of Jung et al. (2020) for an example. Then, we create a sample of $10^6$ simulated $\theta_E$ using the $\rho$ distribution and the $\theta_*$ distribution from the CMD analysis, which yields the probability distribution of $\theta_E$. Here, we only consider the primary lens alone, so $t_{E,i}$ and $t_{E,i}$ are a factor of $\sqrt{1 + q_2 + q_3}$ smaller than the values defined for the triple system.

In Table 7, we summarize the posterior distributions of the physical lens parameters, including the masses of the three lens components, ($M_1$, $M_2$, $M_3$), the projected separation of $M_2$ and $M_3$ to the position of $M_1$, ($r_{\perp,2}$ and $r_{\perp,3}$), the distance to the lens system, $D_L$, and the lens-source relative proper motion, $\mu_{rel}$. We find that the four solutions all consist of a super-Jovian planet in a binary system, but the interpretations of their planetary orbits are different. The ratios of the projected semimajor axes in all four cases are very close to or exceed the conditions for stability (Holman & Wiegert 1999). Of course, the projected semimajor axes are only the minimum separations between the primary and its companions. The stellar companion can lie substantially inside the planet orbit (P-type orbit), or outside of the planet orbit (S-type orbit). As with many Kepler planets orbiting binaries (the first was Doyle et al. 2011, see Martin & Fabrycky 2021 for a complete list), detectability considerations create a bias toward planets near (in the case of Kepler) or appearing to be near (in the case of microlensing) the stability limits (Madsen & Zhu 2019). Thus, the most likely interpretation for the solutions '3L1S A' and '3L1S C', the planet likely orbits the barycentre of a close stellar binary, i.e. a P-type orbit and a circumbinary planet. For the solutions '3L1S B' and '3L1S D', the planet probably orbits the more massive companion of the stellar binary, i.e. a S-type orbit. However, we cannot rule out configurations for any of the four solutions in which the relative locations of the companions are reversed with respect to the primary.

## 5 DISCUSSION

The light curve of the microlensing event OGLE-2019-BLG-1470 shows three distinct features. The first is a smooth bump generated by cusp approach, the other two features originate from a resonant caustic crossing. Our analysis indicates that this event could be explained either by a 3L1S model or by a slightly worse 2L2S model. The 2L2S model is disfavoured by $\Delta\chi^2 \simeq 18$ relative to the best-fitting 3L1S model, its cumulative $\Delta\chi^2$ relative to the best-fitting 3L1S model rises mainly during the caustic crossing region. However, we cannot firmly rule out the 2L2S model with statistical tests. These degenerate models would be resolved if there were high-cadence observations over the peak region.

In this event, the planet manifests itself by generating a resonant caustic which allows forming a detectable anomaly feature when the source passes across the caustic. This is similar to events OGLE-2016-BLG-0613 (Han et al. 2017) and OGLE-2018-BLG-1700 (Han et al. 2020). Planets are still detectable even with small planetary caustics. Actually, this includes a substantial fraction of microlensing planets in binary systems, such as in events OGLE-2006-BLG-284 (Bennett et al. 2020), OGLE-2008-BLG-092 (Poleski et al. 2014), OGLE-2013-BLG-0341 (Gould et al. 2014), and KMT-2019-BLG-1715 (Han et al. 2021a). It seems that the second case (planetary caustics) happens more frequently. However, it is not clear whether this is due to the sample size being small or due to other reasons. On the one hand, the parameter space for generating planetary caustics is larger than that for generating resonant caustics. On the other hand, the sensitivity to planet is higher when there is a resonant caustic. The relative fraction will of course depend on the intrinsic properties of planets in binary systems in the Galaxy. From the theoretical side, it is worth studying the detection probabilities of planets in different configurations of binary systems.

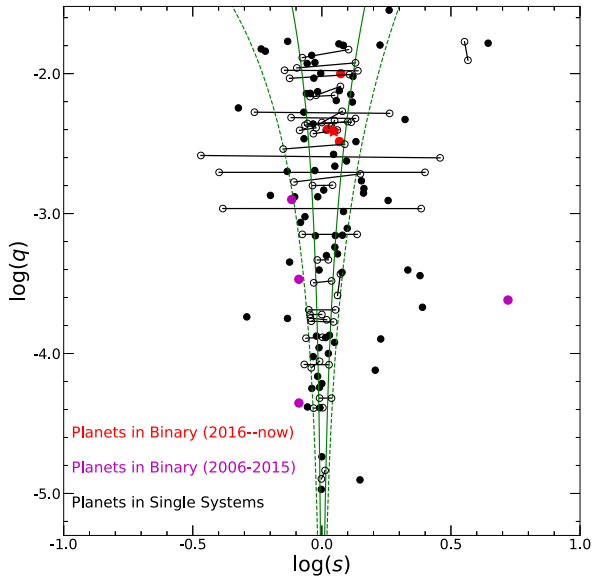

**Figure 12.** (log $s$, log $q$) diagram for microlensing planetary events adapted from fig. 11 of Yee et al. (2021). The black points represent planets around a single star. Magenta and red colours, respectively represent planets in binary systems detected during 2006–2015 and since 2016. The asterisk is the planet found in this paper, and we adopt the parameters of the best-fit 3L1S solution ('3L1S A'). Solutions are considered to be 'unique' (filled points), if there are no competing solutions within $\Delta\chi^2 < 10$. Otherwise, they are shown by pairs of open circles linked by a line segment. 10 planets are excluded because their $q$ differs by more than a factor of two. The two green solid lines represent the boundaries between resonant and non-resonant caustics using equation (59) of Dominik (1999), and the two green dashed lines depict the boundaries for 'near-resonant' caustics proposed by Yee et al. (2021).

Including the candidate lens system, OGLE-2019-BLG-1470LABc, reported in this paper, eight microlensing planets in binary systems have been detected. In Fig. 12, we plot them together with other published microlensing planets in the log $q$ versus log $s$ plane. We separate the eight planets in binary systems into two groups. The magenta points represent the planets detected before 2016, while the red ones are the planets found since 2016. We find that the discovery rate of planets in binary systems has at least doubled since 2016. This is mainly due to the regular operation of KMTNet since 2016 because KMTNet played a major role in all four discoveries since 2016.

However, the two groups exhibit different properties in two aspects: planet-to-host mass ratios and the types of caustic. For the four planets detected before 2016, all of them are located outside the near-resonant range (Yee et al. 2021) with planet-to-host mass ratios log $q < -2.8$. As illustrated by Zang et al. (2021), wide-area high-cadence surveys are sensitive to planets outside the near-resonant range, which is consistent with the detection channels for the four planets in binary systems detected before 2016. Three of them were detected by a pure-survey mode by OGLE and the Microlensing

Observations in Astrophysics (MOA, Sumi et al. 2016) experiments, and OGLE-2007-BLG-349LABc was found by a combination of survey and follow-up. Because KMTNet is more powerful than the previous surveys, it is expected that KMTNet is sensitive to planets out of the near-resonant range for all mass-ratio regions. However, the four planets[8] since 2016 were all detected by a resonant-caustic channel with planet-to-host mass ratios $q \gtrsim 2 \times 10^{-3}$.

This contradiction is similar to the 'missing planetary caustics' problem advocated by Zang et al. (2021), but it is more severe considering the lack of low mass-ratio (log $q < -3$) planets. The four detections before 2016 suggest that low-mass-ratio planets in binary systems are not rare, and the ~20 low-mass-ratio planets in single systems detected by KMTNet have demonstrated its sensitivity to them. Therefore, the problem is likely due to the way we search for planetary signals in binary systems. Indeed, this is proven to be the case for planets orbiting a single star: the advent of systematic KMTNet planetary anomaly searches has started to yield more planets with planetary caustic crossings (Zang et al. 2021, 2022; Gould et al. 2022; Hwang et al. 2022; Wang et al. 2022). These systematic anomaly searches not only increase the total number of known microlensing planets, but also provide complete and homogeneous statistical planetary samples for the studying of, e.g. the planet-to-host mass ratio function.

Previously, the largest such sample was obtained from the wide-area, high-cadence Microlensing Observations in Astrophysics II (MOA-II) survey from 2007 to 2012 (Suzuki et al. 2016). The authors found 23 planets out of 1474 microlensing events with a broken power law for the planet-to-host mass ratio function. Combining planets from two previous studies (Gould et al. 2010; Cassan et al. 2012), they built a sample of 30 planets and found that the power law breaks at mass ratio $q_{\rm br} \equiv 1.7 \times 10^{-4}$, i.e. cold Neptunes are likely the most common type of planets beyond the snow line (for late dwarfs). Furthermore, a statistical work based on long period ($\gtrsim 2$ yr) transiting planet candidates from the prime Kepler mission found a compatible result that the long-period Neptune-sized planets are at least as common as the Jupiter-sized planets (for FGK dwarfs, Kawahara & Masuda 2019). The authors pointed out that it is essential to quantify the completeness of smaller planets to facilitate more detailed comparisons.

We note that there are two microlensing events of planet in binary system during 2007–2012. The first is OGLE-2007-BLG-349 (MOA-2007-BLG-379, Bennett et al. 2016), which contains a planet with mass ratio $q \simeq 3.4 \times 10^{-4}$. This planet was included as one of the 23 planets in the statistical study of Suzuki et al. (2016).[9] The second event, OGLE-2008-BLG-092 (Poleski et al. 2014), is not a MOA event. The fraction of planets in binary systems appears to be low (1/23) in the sample of Suzuki et al. (2016).

However, as more planets in binary systems are discovered, one may need to be cautious about whether these planets can be included in such statistical studies. For the KMTNet data, the AnomalyFinder algorithm is efficient in uncovering the buried planetary signals (Zang et al. 2021), including signals from low-mass-ratio ($q \lesssim 10^{-4}$) planets (Hwang et al. 2022). But the AnomalyFinder algorithm has not been applied to KMTNet binary events to find potential planetary anomalies. The lack of planetary caustics in the current KMTNet sample of planets in binary systems indicates that this sample may be incomplete. Thus, these planets (in binaries) cannot be included in the current study of KMTNet mass-ratio function. Systematic KMTNet planetary anomaly searches for planets in binary systems are therefore needed. With the successful implementation of the current KMTNet AnomalyFinder algorithm, one may naturally think of applying this algorithm to the residuals of 2L1S light curves. However, this approach requires a significant effort on careful 2L1S modelling (about 200 events per year) with the inclusion of high-order effects (parallax and orbital motion of the stellar binary) in many cases, as well as substantial additional work for data reductions. On the other hand, the reward is also rich because it will give, for the first time, the statistics of microlensing planets in binary systems. We plan to pursue this in the future.

---

[8]There is one more case if we count the event OGLE-2019-BLG-0304 (Han et al. 2021b) as a candidate of planet in binary system. The 3L1S model of OGLE-2019-BLG-0304 includes a planet with planet-to-host mass ratio $q = 1.82 \pm 0.26 \times 10^{-3}$. The planet also generates a resonant caustic.

[9]This planet was also included as one of the six planets used to statistically investigate the frequency of solar-like systems and of ice and gas giants (Gould et al. 2010). The authors have realized that the OGLE-2007-BLG-349 system contains a third body, but difficult to fully characterize at that time.

## ACKNOWLEDGEMENTS

We thank the referee Daniel Bramich for a critical report that improved the paper significantly. R.K., W.Zang, H.Y., S.M., and W.Zhu acknowledge support by the National Science Foundation of China (Grant No. 12133005). This research has made use of the KMTNet system operated by the Korea Astronomy and Space Science Institute (KASI) and the data were obtained at three host sites of CTIO in Chile, SAAO in South Africa, and SSO in Australia. Work by C.H. was supported by the grants of National Research Foundation of Korea (2019R1A2C2085965 and 2020R1A4A2002885). J.C.Y. acknowledges support from N.S.F Grant No. AST-2108414. W.Zhu acknowledges the science research grants from the China Manned Space Project with No. CMS-CSST-2021-A11. Y.S. acknowledges support from BSF Grant No. 2020740. The authors acknowledge the Tsinghua Astrophysics High-Performance Computing platform at Tsinghua University for providing computational and data storage resources that have contributed to the research results reported within this paper. This research has made use of the NASA Exoplanet Archive, which is operated by the California Institute of Technology, under contract with the National Aeronautics and Space Administration under the Exoplanet Exploration Program.

## DATA AVAILABILITY

The data underlying this article will be shared on reasonable request to the corresponding author.

## APPENDIX A: COMBINING TWO 2L1S MODELS TO FORM A 3L1S MODEL

The magnification pattern produced by a triple-lens system is complex and difficult to calculate. Previous studies showed that the magnification of a triple-lens system which contains two planets can be calculated as the summation of the magnifications from two binary-lens systems (binary superposition). This is valid for high magnification cases (Rattenbury et al. 2002; Ryu, Chang & Park 2011) as well as for both planetary caustics (Han et al. 2001) and central caustics (Han 2005).

For the case of a planet in a binary-star system, the superposition of magnifications is no longer valid. In Han et al. (2001), they found that if the heavier companion mass ratio $\gtrsim 0.05$, the magnification deviation from the binary superposition becomes considerable. An intuitive explanation is that the caustics produced by the planet will be easily affected by the binary-star system. Instead of seeking for valid superposition for magnification calculations, we expect that in some cases, topologically the overall caustic structure corresponding to the planet remains the same after we add the planet component to a binary-star model. In this case, the binary superposition is 'valid' in the sense that the caustic structure required to produce all anomaly features in the light curve still exist. The resulting triple-lens model can be taken as an initial approximation for more accurate modelling.

Now we investigate how to combine two binary-lens models to form an initial triple-lens model. Specifically, we focus on combining a binary-star model and a planetary model. Readers who are not interested in the technical details can directly go to appendix A3 which gives a short recipe for the procedures.

We denote the parameters of the binary-star model and the planetary model as $(t_{0, \mathrm{B}}, u_{0, \mathrm{B}}, t_{\mathrm{E, B}}, \rho_{\mathrm{B}}, s_{\mathrm{B}}, q_{\mathrm{B}}, \alpha_{\mathrm{B}})$ and $(t_{0, \mathrm{c}}, u_{0, \mathrm{c}}, t_{\mathrm{E, c}}, \rho_{\mathrm{c}}, s_{\mathrm{c}}, q_{\mathrm{c}}, \alpha_{\mathrm{c}})$, respectively.[10] The goal is to obtain the parameters $(t_0, u_0, t_{\mathrm{E}}, \rho, s_2, q_2, s_3, q_3, \alpha, \psi)$ of the triple-lens system. For parameters other than $(s_3, q_3, \psi)$, we use the same values as the binary-star system:

$$(t_0, u_0, t_{\mathrm{E}}, \rho, s_2, q_2, \alpha) = (t_{0,\mathrm{B}}, u_{0,\mathrm{B}}, t_{\mathrm{E,B}}, \rho_{\mathrm{B}}, s_{\mathrm{B}}, q_{\mathrm{B}}, \alpha_{\mathrm{B}}), \quad (A1)$$

since the extra planet would not change these parameters significantly. We derive the remaining parameters, i.e. $(s_3, q_3, \psi)$ from the two binary-lens models.

We designate the masses of the three lens objects as $M_1, M_2,$ and $M_3$ (with $M_1 > M_2 > M_3$ and $\Sigma_{i=1}^{3} M_i = 1$). We use the same coordinate system as in Kuang et al. (2021). $M_1$ and $M_2$ are located along the horizontal axis, and their centre of mass is the origin. Specifically, their masses and positions $(x_i, y_i)$ are:

$$M_1 = 1/(1 + q_2 + q_3), \quad M_2 = q_2 M_1, \qquad M_3 = q_3 M_1,$$
$$x_1 = -q_2 s_2/(1 + q_2), \qquad y_1 = 0,$$
$$x_2 = s_2/(1 + q_2), \qquad y_2 = 0,$$
$$x_3 = x_1 + s_3 \cos \psi, \qquad y_3 = y_1 + s_3 \sin \psi. \quad (A2)$$

---

[10]We note that the $(t_{\mathrm{E}}, \rho)$ will not be the same from two individual 2L1S modelling with two different data sets, such as the models shown in Table 3. One can obtain two 2L1S models with the same $(t_{\mathrm{E}}, \rho)$ with a joint fit.

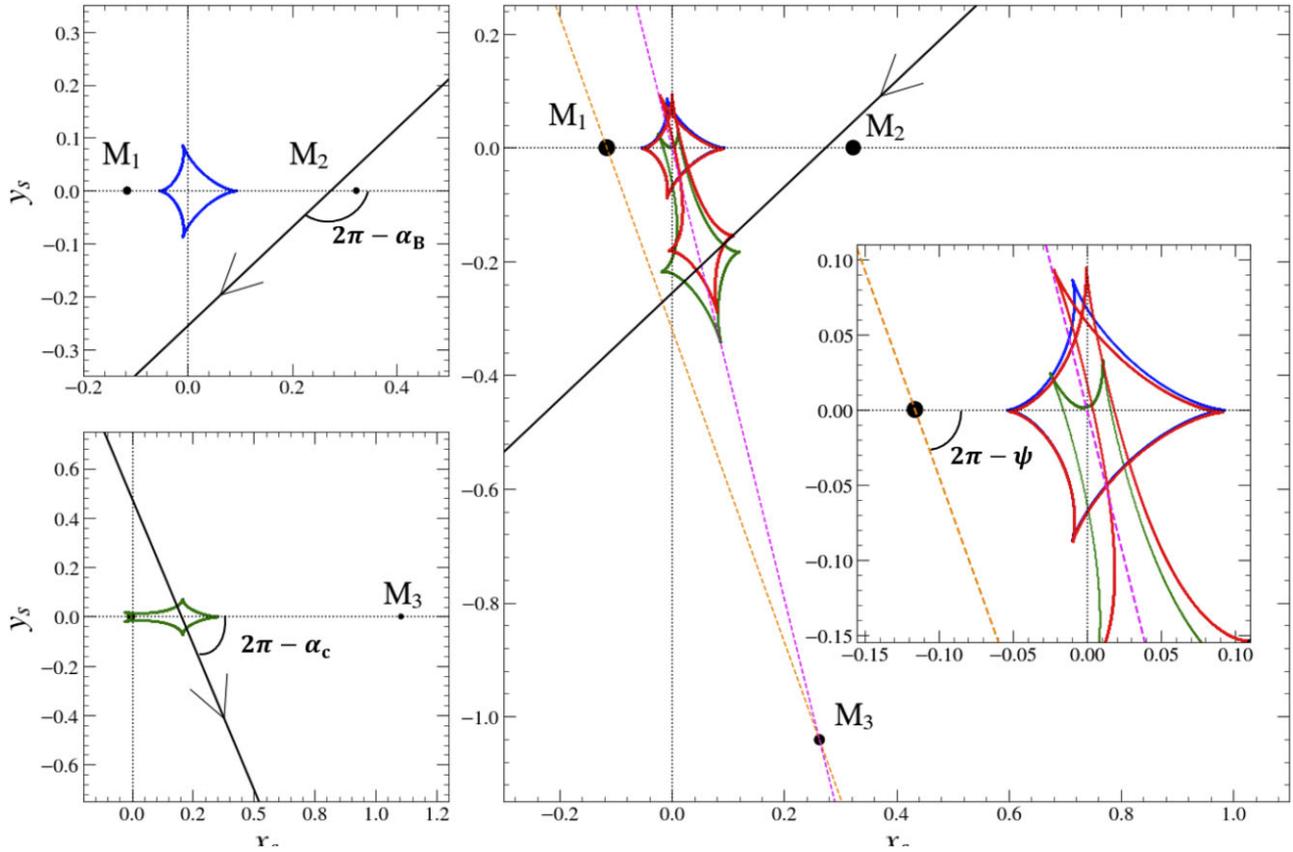

**Figure A1.** An illustration of combining two binary-lens models to form a triple-lens model. In the upper left-hand panel, we show the first binary-lens model $M_1$–$M_2$ with parameters $(s_B, q_B, \alpha_B) = (0.439, 0.359, 223°)$, produces a diamond-shaped caustic (in blue) near the origin. The lower left-hand panel shows the second binary-lens model with parameters $(s_c, q_c, \alpha_c) = (1.108, 3.472 \times 10^{-3}, 293°)$, which produces a resonant caustic (in green). The right-hand panel shows the resulting triple-lens model constructed by combining the two binary-lens models in the left-hand panel. The red curve shows the caustics of the triple-lens model. The blue curve is the same caustic as in the upper left-hand panel, the green curve is plotted by rotating by an angle so that it lies on the line (dashed magenta) connecting $M_3$ and the origin. The black dots are the lenses. The black solid lines with arrows show the source trajectories.

We show two example cases, corresponding to close ($s_B < 1$) and wide ($s_B > 1$) binary-star system, respectively.

### A1 Close binary-star system ($s_B < 1$)

For the first case, $(s_c, q_c)$ of the planetary lens system can be taken as $(s_3, q_3)$, and $\psi$ calculated by subtracting $\alpha_c$ from $\alpha_B$,

$$s_3 = s_c,$$
$$q_3 = q_c,$$
$$\psi = \alpha_B - \alpha_c. \tag{A3}$$

Fig. A1 shows the related geometries. The upper left-hand panel shows the caustics (in blue) of the binary-star model. The lower left-hand panel shows the caustics (in green) of the planetary model. In the right-hand panel, the red curve is the caustics produced by the resulting triple-lens model. The overall caustic structure remain the same as the two individual binary-lens models. So, the combined triple-lens model can be taken as an initial model for further optimising.

We note that the caustic corresponding to $M_3$ is not along the line connecting $M_3$ and $M_1$ (the orange dashed line), but close to a line (the magenta dashed line) connecting $M_3$ and the origin, i.e. the centre of mass of $M_1$–$M_2$. So, one can regard $M_1$–$M_2$ as a whole and has a net effect on $M_3$. In this case, the position and planet-to-host mass ratio of the planet, parametrized with $(s_c, q_c, \psi)$ are relative to the 'effective' mass of $M_1$–$M_2$, instead of being relative only to $M_1$. The extreme case is that $s_B \to 0$, i.e. $M_1$ and $M_2$ are merged into one object.

The effective lensing position of a component in the binary lens system is shifted toward its companion (Di Stefano & Mao 1996; An & Han 2002). A lens component $i$ will shift toward its companion $j$ by an amount (Chung et al. 2005):

$$\Delta x_{L,i \to j} \simeq \frac{M_j/M_i}{(s + s^{-1})/(\theta_{E,i}/\theta_E)} \frac{\theta_{E,i}}{\theta_E}, \tag{A4}$$

where $M_i$, $M_j$ are the masses of the individual lens components, $\theta_{E,i}$ and $\theta_E$ are the Einstein ring radius corresponding to $M_i$ and $M_i + M_j$, respectively. In our case, the effective lensing position of $M_1$ will shift toward $M_2$ by an amount of

$$\Delta x_{1 \to 2} \simeq \frac{q_B}{(s_B + s_B^{-1})(1 + q_B)} \approx \frac{s_B q_B}{1 + q_B}, \tag{A5}$$

i.e. for the case of $s_B < 1$, the effective lensing position of $M_1$ is close to the centre of mass of $M_1$ and $M_2$. If we regard $M_1$ and $M_2$ are effectively located at the origin with a mass $M_1 + M_2$, then we have the following relations,

$$M_2/M_1 = q_B, \qquad M_3/(M_1 + M_2) = q_c, \qquad \Sigma_{i=1}^3 M_i = 1,$$
$$\psi_0 = \alpha_B - \alpha_c,$$
$$x_3 = s_c \cos \psi_0, \qquad y_3 = s_c \sin \psi_0,$$

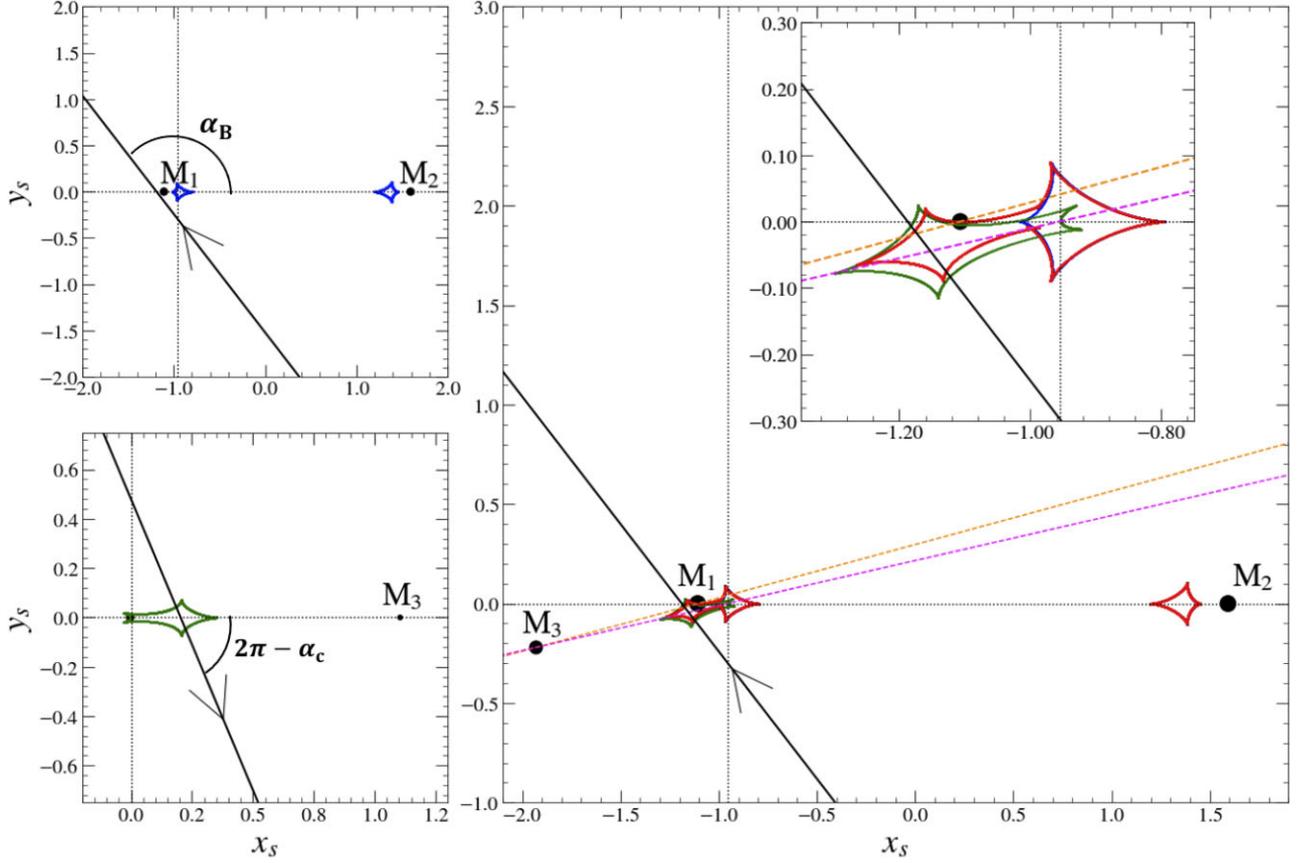

**Figure A2.** The same as Fig. A1. The binary-star $M_1$–$M_2$ are in wide orbit that $(s_B, q_B, \alpha_B) = (2.700, 0.695, 128°)$. The planetary lens system in the bottom left-hand panel is the same as in Fig. A1. The green curve in the right-hand panel is plotted by rotating an angle and then shifting leftwards so that it lies in the line connecting $M_3$ and the geometric centre (calculate numerically, $x \approx -0.955$) of the diamond-shaped caustic near $M_1$.

(A6)

where $\psi_0 \equiv \angle M_2 O M_3$, $O$ is the origin. We can derive $(s_3, q_3, \psi)$ to match our coordinate definition as

$$q_3 = q_c(1 + q_B),$$
$$s_3 = \sqrt{(x_3 - x_1)^2 + (y_3 - y_1)^2},$$
$$\psi = \arctan((y_3 - y_1)/(x_3 - x_1)).$$

(A7)

Equations (A3) and (A7) are two different ways of adding a planetary mass component into an existing binary-star model. For a close binary system, the resulting caustics have similar structure. So both equations are valid for obtaining an initial triple-lens model.

### A2 Wide binary-star system ($s_B > 1$)

For the second case, $q_c$ of the planetary model can be taken as $q_3$. However, we cannot simply take $s_c$ as $s_3$. This is because $M_2$ has smaller effect on $M_3$ when $s_B > 1$. The extreme case is that $s_B \to +\infty$. In the planetary 2L1S model, the total mass $\sim M_1 + M_3$, while in the resultant 3L1S model, the total mass equals to $M_1 + M_2 + M_3$. So they correspond to different $\theta_E$'s. So considering $s_c$ and $s_3$ are in

units of different $\theta_E$'s, we have

$$s_3 = s_c \sqrt{(1 + q_c)/(1 + q_c + q_B)},$$
$$q_3 = q_c,$$
$$\psi = \alpha_B - \alpha_c.$$

(A8)

In this way we can retain the caustic structure produced by the planet, as shown in Fig. A2. Similar to the previous case in appendix A1, the caustic corresponding to $M_3$ is also not along the line connecting $M_3$ and $M_1$ (the orange dashed line). Instead, the caustic lies close to the line (the magenta dashed line) connecting $M_3$ and the geometric centre of the diamond-shaped caustic (the one close to $M_1$). The position of this geometric centre (the mean values of the coordinates of all points at that caustic) is calculated numerically after we obtained the caustic shown in the upper left-hand panel. We can also estimate the effective lensing position of $M_1$ in this case, as in equation (A5),

$$\Delta x_{1 \to 2} \simeq \frac{q_B}{(s_B + s_B^{-1})(1 + q_B)} \approx \frac{q_B}{s_B(1 + q_B)}.$$

(A9)

For the example case shown in Fig. A2, $(s_B, q_B) = (2.700, 0.695)$, the effective lensing position of $M_1$ is:

$$x_1 + \Delta x_{1 \to 2} = -\frac{q_B s_B}{1 + q_B} + \frac{q_B}{s_B(1 + q_B)} \approx -0.955,$$

(A10)

which is exactly equal to the position of the numerically calculated geometric centre of the diamond-shaped caustic near $M_1$.

In summary, for the above two examples, the effective lensing position of $M_1$ will shift towards $M_2$ by

$$\Delta x_{1\rightarrow 2} = \begin{cases} \frac{q_B}{1+q_B}s_B, & s_B \lesssim 1. \\ \frac{q_B}{1+q_B}s_B^{-1}, & s_B \gtrsim 1. \end{cases} \quad (A11)$$

The above equations may not be applied for other values of $(s_B, q_B)$. For example, the separation at which the binary-lens system can be regarded as 'wide binary' (when there are isolated regions of magnification) depends on the mass ratio between the binary components (Di Stefano & Mao 1996). Besides, in triple-lens system, the location and mass of the third lens ($M_3$) need to be taken into account. We leave the determination of the effective lensing position and mass of the $M_1$–$M_2$ binary system, as seen from $M_3$, as a future work.

## A3 Recipe

In Appendixes A1–A2, we give detailed justifications for how to combine two 2L1S models. Here we give a short recipe of the procedures to obtain an initial triple-lens model applicable to the case where a light curve shows distinct anomalies from a binary system and a planetary system as in this event, OGLE-2019-BLG-1470.

(i) Excluding part of the data points in corresponding anomaly (planetary or binary) regions. Obtaining two different data subsets.

(ii) Modelling with 2L1S model for these two data subsets. Obtaining two sets of 2L1S model parameters, i.e. $(t_{0,B}, u_{0,B}, t_{E,B}, \rho_B, s_B, q_B, \alpha_B)$ for the binary-star system and $(t_{0,c}, u_{0,c}, t_{E,c}, \rho_c, s_c, q_c, \alpha_c)$ for the planetary system.

(iii) Combining these two sets of 2L1S model parameters to form the initial parameters $(t_0, u_0, t_E, \rho, s_2, q_2, s_3, q_3, \alpha, \psi)$ of the triple-lens model which will retain the required caustic structure. For the case of close binary ($s_B < 1$), the parameters of the initial triple-lens model can be obtained from equations (A2) and (A3) (or A7). While for the case of wide binary ($s_B > 1$), one can use equations (A2) and (A8).

(iv) Finally, in both example cases, the planet has little effect on the caustics produced by the binary star. On the other hand, the caustics produced by the planet can be easily affected by the binary-star system. After obtaining the rough parameters for a triple model and before further optimisation, one may need to manually fine-tune the triple-lens parameters (mainly on $s_3, q_3, \psi$) to adjust the caustics to the desired position.

[1]*Department of Astronomy, Tsinghua University, Beijing 100084, China*
[2]*Department of Engineering Physics, Tsinghua University, Beijing 100084, China*
[3]*Korea Astronomy and Space Science Institute, Daejeon 34055, Republic of Korea*
[4]*University of Science and Technology, 217 Gajeong-ro Yuseong-gu, Daejeon 34113, Republic of Korea*
[5]*Astronomical Observatory, University of Warsaw, Al. Ujazdowskie 4, PL-00-478 Warszawa, Poland*
[6]*National Astronomical Observatories, Chinese Academy of Sciences, Beijing 100101, China*
[7]*University of Canterbury, Department of Physics and Astronomy, Private Bag 4800, Christchurch 8020, New Zealand*
[8]*Max-Planck-Institute for Astronomy, Königstuhl 17, D-69117 Heidelberg, Germany*
[9]*Department of Astronomy, Ohio State University, 140 W. 18th Ave., Columbus, OH 43210, USA*
[10]*Department of Physics, Chungbuk National University, Cheongju 28644, Republic of Korea*
[11]*Department of Particle Physics and Astrophysics, Weizmann Institute of Science, Rehovot 76100, Israel*
[12]*Center for Astrophysics | Harvard, Smithsonian, 60 Garden St., Cambridge, MA 02138, USA*
[13]*School of Space Research, Kyung Hee University, Yongin, Kyeonggi 17104, Republic of Korea*
[14]*Department of Physics, University of Warwick, Gibbet Hill Road, Coventry, CV4 7AL, UK*
[15]*Harvard College, Harvard University, MA 02138, USA*

This paper has been typeset from a TEX/LATEX file prepared by the author.